\documentclass[final,1p,times]{elsarticle}
\usepackage{graphicx}
\usepackage{amssymb}

\usepackage{lineno}
\usepackage[utf8]{inputenc}
\usepackage[T1]{fontenc}

\usepackage{graphicx}
\usepackage{dcolumn}
\usepackage{bm}
\usepackage{framed}

\newcommand{\Ig}{\hbox{$U$}} 
\newcommand{\Sp}{\hbox{$I$}} 
\newcommand{\St}{\hbox{$R$}} 
\newcommand{\SIR}{\hbox{$SIR$}}

\newcommand{\Inf}{\hbox{$\sigma$}}

\newcommand{\Sd}{\hbox{\textbf{S}}}

\newcommand{\NVI}{\hbox{$NVI$}}
\newcommand{\setX}{\hbox{$\mathbf{X}$}}
\newcommand{\setY}{\hbox{$\mathbf{Y}$}}

\newcommand{\dSt}[1]{\hbox {$\widehat{\Inf}_{#1}$}}

\newcommand{\NC}{\emph{NA}}
\newcommand{\AN}{\emph{AN}}
\newcommand{\DN}{\emph{DN}}

\def\mathbi#1{\textbf{\em #1}}

\usepackage{colortbl}
\usepackage{color}
\definecolor{Gray}{gray}{0.87}
\usepackage{subfigure}

\begin{document}

\begin{frontmatter}

\title{Influence maximization by rumor spreading on correlated networks through community identification}

\author[mainaddress,secondmainaddress]{Didier A. Vega-Oliveros\corref{mycorrespondingauthor}}
\cortext[mycorrespondingauthor]{Corresponding author}
\ead{davo@icmc.usp.br}

\address[mainaddress]{Departamento de Computa\c{c}\~{a}o e Matem\'{a}ticas, \\
Faculdade de Filosofia Ci\^{e}ncias e Letras de Ribeir\~{a}o Preto,
Universidade de S\~{a}o Paulo -- Ribeir\~{a}o Preto, SP, Brazil.}

\address[secondmainaddress]{School of Informatics, Computing and Engineering, Indiana University -- Bloomington, IN, USA.}

\author[secondaryaddress]{Luciano da Fontoura Costa}
\address[secondaryaddress]{Instituto de F\'{i}sica de S\~{a}o Carlos, 
Universidade de S\~{a}o Paulo -- S\~{a}o Carlos, SP, Brazil.}

\author[thirdaddress]{Francisco Aparecido Rodrigues}
\ead{francisco@icmc.usp.br}
\address[thirdaddress]{Departamento de Matem\'{a}tica Aplicada e Estat\'{i}stica,\\
Instituto de Ci\^{e}ncias Matem\'{a}ticas e de Computa\c{c}\~{a}o,
Universidade de S\~{a}o Paulo -- S\~{a}o Carlos, SP, Brazil.}


\begin{abstract}
The identification of the minimal set of nodes that maximizes the propagation of information is one of the most relevant problems in network science. In this paper, we introduce a new method to find the set of initial spreaders to maximize the information propagation in complex networks. We evaluate this method in assortative networks and verify that degree-degree correlation plays a fundamental role in the spreading dynamics. Simulation results show that our algorithm is statistically similar, regarding the average size of outbreaks, to the greedy approach in real-world networks. However, our method is much less time consuming than the greedy algorithm. 
\end{abstract}


\end{frontmatter}


\section{\label{sec:Introduction}Introduction}

With the popularization of Internet access by mobile devices, online social networks have emerged as a suitable medium for information transmission~\cite{Kempe15,Pastor-Satorras2015}. News, rumors, and advertisements propagate fast in these networks due to the low average degree of separation between users~\cite{Pastor-Satorras2015}. Information is also exchanged in communication networks, where users share files related to multiple contents, including images, audio, and video. Communication and social networks are also characterized by a very heterogeneous structure, in which most of the users are low connected, whereas a minimal set of them have many connections~\cite{Pastor-Satorras2015}. Moreover, in some social networks, high degree vertices tend to connect to low degree vertices, defining a disassortative wiring pattern. This complex structure of networks affects the information propagation, defining a hierarchy among the nodes~\cite{Kempe15}.  This means that networks present special nodes that are the most influential spreaders in the propagation process~\cite{morone2015influence, Vega-OliverosH17}, i.e., nodes that maximize the average size of outbreaks.

The identification of these influential nodes is essential to understand and control the spreading process on social networks~\cite{morone2015influence}. Particularly, the influence maximization problem (IMP) is faced with the selection of a set of $\eta$ spreaders that trigger the largest cascade of new adopters according to a spreading dynamic~\cite{Richardson2002}. The problem of finding this set of initial spreaders is NP-hard for most of the spreading models~\cite{Kempe15}, which makes the IMP as a challenge for network scientists. Thus, since it is not possible to obtain the optimal results for most of the networks, the IMP is addressed by heuristic algorithms. For instance, one of the most studied methods is a hill-climbing greedy approach~\cite{Kempe15}, which guarantees that the influence spread is within $(1 - 1/e)$ of the optimal influence spread. This greedy algorithm outperforms the classic degree and centrality-based heuristics in influence spread~\cite{Kempe15}, but it is still very computationally expensive. In addition, Morone and Makse~\cite{morone2015influence} mapped the IMP onto optimal percolation in random networks to identify the nodes that should be removed to minimize the average size of outbreaks. They verified that this set is given by the nodes whose removal break down the network into many disconnected subgraphs. However, this set of nodes does not correspond necessarily to optimal spreaders, as verified by Radicchi and Castellano~\cite{Radicchi07}. Although all these works advanced the study of influence maximization, they disregard patterns of connections, such as degree-degree correlation and community structure, which have a fundamental impact on spreading dynamics~\cite{Pastor-Satorras2015}. 

Degree-degree correlations (or assortativity) is a network property in which nodes with similar features, such as degree, tend to be connected. Previous works verified that epidemics spread faster in assortative networks, but the reach is more extensive on disassortative structures~\cite{kiss2008effect}. Assortativity also influences the spreading threshold~\cite{Boguna2003} and the diffusion time~\cite{Bertotti2016}. Although degree-degree correlation influences the spreading dynamics, the role of this network property on the influence maximization problem has not been addressed yet  (see, for instance, \cite{Kempe15,Pastor-Satorras2015}). Here, we analyze how degree-degree correlation affects the average size of outbreaks in rumor dynamics. 

We also propose a method for identification of the most influential spreaders based on community organization. Communities are groups of nodes densely connected among them, but with few connections with other groups~\cite{Fortunato16}. Some authors verified that to improve the spreading efficiency, a good strategy is to distribute the seeds on the network producing lower overlap~\cite{BalkanskiIS17,Galstyan2009,Wang2010,Weng2013}. If the community structure is not considered, then only suboptimal solutions can be obtained~\cite{Galstyan2009}. This happens because vertices belonging to the same community are likely to be more similar to each other and share the same set of neighbors. Although communities influence the diffusion of information, only a few studies have considered  the community organization to study the influence maximization problem 
~\cite{Galstyan2009, Wang2010, Weng2013, Cao2011, Zhang2013, Vega2015spreader, Hosseini-Pozveh2016}. Indeed, most of these works try to reduce the number of candidate vertices according to some evaluation method and the community structure. For instance, Galstyan et al.~\cite{Galstyan2009} employed the greedy approach for selecting the seeds in the smallest community and verified that this might cause a global activation cascade even for a small number of seeds.  However, the results are restricted to random networks made up of two communities. Wang et al.~\cite{Wang2010} introduced a community-based greedy algorithm to find the $\eta$ most influential nodes. The idea is to divide the network into communities and then, by a dynamic programming algorithm, incrementally select the community from which the next influential node is taken. The method involves high computational cost, although it is an order of magnitude faster than the greedy algorithm. In a similar approach, Cao et al.,~\cite{Cao2011} transformed the influence maximization problem into an optimal resource allocation problem in the network communities. Initially, the method assumes that the communities are disconnected. Then, the method selects $\eta$ candidates from each community according to the degree centrality and a dynamic programming algorithm identifies the final target nodes. 

Although these works provided essential results on the influence maximization problem, none of them addressed the impact of the assortativity on the propagation dynamics. These methods are computationally expensive and consider a relatively low number of initial spreaders, i.e., up to $\eta = 50$ spreaders. Moreover, classical rumor models are not addressed by these studies, although the model by Maki and Thompson~\cite{Maki1973} is often used to study information dynamics in networks~\cite{Pastor-Satorras2015,Vega-Oliveros2019a, Zanette2001a,moreno2004dynamics, Borge-Holthoefer2012}. Thus, in the present work, we provide an analysis of the impact of degree-degree correlation on the influence maximization problem, where the Maki-Thompson algorithm models the information spreading. A simple approach to maximize information diffusion considering the community structure of the network is introduced. We perform exhaustive simulations in eight real and ten artificial complex networks and verify that assortativity plays a significant role in the influence maximization problem. For instance, increasing the number of initial spreaders may not increase the size of the outbreak. Moreover, the selection of influential spreaders through communities is statistically similar to the greedy algorithm. However, our method requires much lower computational cost and, therefore, is more suitable in practice.

\section{\label{sec:Method}Concepts and methods}

A social network can be represented as a graph $G = (V, E)$ made up of a set of $N = |V|$ vertices (nodes) and a set $E$ of edges that connect pairs of vertices. Here, we consider only undirected and static networks.  The degree $k_i$ of a vertex $i$ corresponds to the number of edges attached to $i$. The degree distribution of a network $P(k)$ gives the probability that a given randomly selected vertex has degree $k$. Social networks are characterized by highly heterogeneous degree distribution, presenting a scale-free organization~\cite{Pastor-Satorras2015}, where most of the nodes are low connected, but a small set of nodes have a high degree. We can also analyze the connection pattern of vertices with the degree-degree correlation. In assortative, or positively correlated, networks nodes of similar degree tend to be connected. In disassortative, or negatively correlated, networks low-degree nodes tend to connect with strongly connected vertices. If the tendency of connection is independent of the node degree, then the network is called non-assortative. The level of assortativity can be quantified by the Pearson correlation coefficient, $\rho$, of the degrees of nodes at either end of an edge~\cite{Newman2002}. According to this measure, a network can be classified as (i) assortative ($\rho > 0$), (ii) disassortative ($\rho < 0$), or (iii) non-assortative ($\rho \approx 0$). Degree-degree correlation plays a fundamental role in the analysis of several dynamical processes in networks~\cite{EstradaAssorta2011}, like network evolution and link prediction~\cite{Vega-Oliveros2019a},  epidemic spreading~\cite{Newman2002}, or synchronization~\cite{Peron015}.

\subsection{Influence models}

One can approach the spreading of rumors or information as a psychological contagion where an idea ``contaminates'' the mind of a population~\citep{Pastor-Satorras2015,Vega-Oliveros2019a}. In the general rumor approach~\cite{Maki1973,moreno2004dynamics,Vega-Oliveros2019a} ignorant nodes (\Ig) are those who are unaware of the information, spreaders (\Sp) are informed individuals that transmit the rumor, and stiflers (\St) are individuals who have heard the rumor but do not spread the information anymore~\cite{Pastor-Satorras2015, Vega-OliverosH17}. Thus, each subject can be in one of the three states, i.e., unaware, spreader, or stifler, at each time step. Notice that stiflers act as recovered individuals in a disease spreading model, as they do not participate in the spreading process anymore~\cite{Maki1973, Pastor-Satorras2015}. Rumor models are different from the traditional susceptible-infected-recovered (\SIR) spreading model, in the sense that the spreading of information is intentional and the transition between states occurs only through contacts, whereas the transition from infected to recovery in the {\SIR} model occurs spontaneously, independent of the connections. In other words, {\SIR} models assume that people have the same behavior to stop spreading information.

In the rumor model proposed by Maki and Thompson~\cite{Maki1973}, a node that knows the rumor tries to pass the information to each of its neighbors according to a probability $\beta$. When the contact is performed between two informed individuals, the active spreader becomes a stifler according to probability $\mu$. We consider this model because all transitions occur through contacts, which makes the dynamics strongly dependent on the network structure. In addition, this model simulates real social dynamics more accurate than epidemic or threshold models, as people may have multiple opinions, i.e., positive, hesitating or negative~\cite{StieglitzD13}. In rumor propagation, spreaders behave like individuals with positive views and stiflers with negative ones.

Different from information cascade and threshold models~\cite{Kempe15,Pastor-Satorras2015}, spreaders stop the propagation after considering the information has lost its ``news value''~\cite{Maki1973}, i.e., the rumor is broadly known or without novelty according to the contact interaction. Besides, the psychological contagion implies a bigger exposure in time and different sources of information. Then, spreaders in rumor models  try repeatedly to propagate the information to their neighbors, which is related to a social reinforcement action.

We set $\beta = 0.3$ and $\mu = 0.2$ for all simulations. However, and without loss of generality, the results are stable in a larger combination of spreading parameters, as shown in \ref{sec:appendixB}. With the selection of this high ratio, we assure to be in a regime above the critical point of the spreading process for all the considered networks. Opposite to epidemics, where the efforts are in restrain the propagation, here we are interested in analyzing the behavior of influence maximization in a ``pandemic'' or viral marketing scenario~\cite{Hu_2018}, far from the critical regime scenario.

\subsection{Influence maximization}

The propagation impact ($\Inf({\Sd})$) for the set {\Sd}  of seeds corresponds to the expected fraction of vertices that were informed during the spreading process. Therefore,
the influence maximization problem (IMP) seeks the set {\Sd} of vertices that contain $|\Sd| = \eta$ initial seeds and maximize the reach of information, i.e., $\Inf({\Sd})$ should be maximized as a function of the set {\Sd}.

Let us consider a discrete diffusion scenario in which each vertex $i$ can be in only one state at each time step. The initial conditions for the influence maximization problem is defined as $\Ig(0)  =  V \, \backslash \, \Sd$, $\Sp(0) \, = \, \Sd $ and $\: \St(0) \, = \, \{\emptyset\} $, where $\Ig(0)$ represents the set of unaware individuals in time $t=0$ and $\Sp(0)$ the set of initial spreaders or initial seeds \Sd . At each time step, all spreaders uniformly try to infect their neighbors with probability $\beta$, or stop the diffusion with probability $\mu$ according to the truncated dynamics~\cite{Borge-Holthoefer2012}. More specifically, a spreader tries to inform each of its neighbor until meets another informed node and becomes a stifler. The process ends when $\Sp(\infty) = \{\emptyset\}$ and then we can calculate the final fraction of informed individuals  ($\dSt{\Sd}$), i.e. $\dSt{\Sd} = {|\St(\infty)|}/{N}$ or $\dSt{\Sd} =  1 - {|\Ig(\infty)|}/{N}$. However, $\dSt{\Sd}$ is a function with stochastic fluctuations. Thus, the influence function, $\Inf(\Sd)$, is estimated by performing a sufficient number of calculations of the final fraction of informed individuals $\dSt{\Sd}$, with the set of initial spreaders $\Sd$:
\begin{equation}
\Inf(\Sd) = \frac{1}{K}\sum_{\kappa = 1}^{K}{\dSt{\Sd}}^{\kappa}
\label{eq:infFunc}
\end{equation} where ${\dSt{\Sd}}^{\kappa}$ represents the final fraction of informed individuals for a particular run $\kappa$, and $K$ is the total number of simulations in order to obtain a good estimate of the mean value of $\Inf(\Sd)$.
Here, we define the total number of simulations for each set of initial nodes as $K = 600$. This choice is due to the computational constraint of the Greedy (hill-climbing) method, which has a very high computation cost.  However, we verify that the results are stable for larger values of $K$ in artificial networks.

\begin{figure}[!tb]
\centering
\includegraphics[scale=0.3]{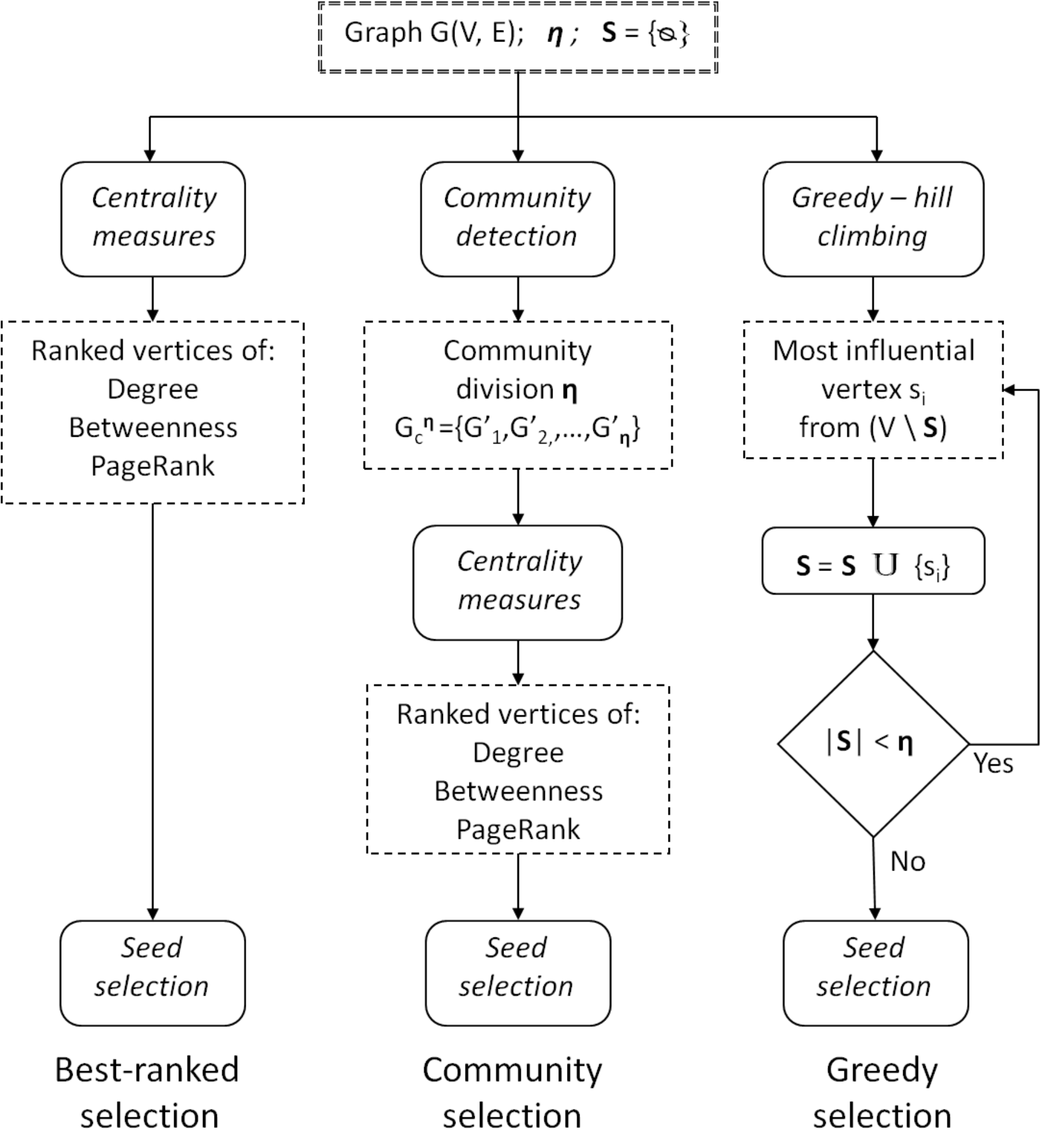}     
\caption{\label{fig:seedsProp} Methods considered here to address the influence maximization problem (IMP).}
\label{fig:schema}
\end{figure}

Fig.~\ref{fig:schema} shows the three methods considered here for solving the influence maximization problem: (\textit{i}) by selecting the $\eta$ vertices with the highest value of a given centrality measure; (\textit{ii}) by detecting the $\eta$ communities on the network and selecting the most central nodes inside each community; and (\textit{iii}) by a greedy approach, that is a hill-climbing procedure that returns the $\eta$ most influential nodes. These methods  receive as input the network $G$ and return a set of $\eta$ initial spreaders.

Methods based on network centrality assume that the most central nodes convince the largest number of individuals on the network~\cite{Kitsak2010,Kandhway2016}. However, the problem, in this case, is how to select the most suitable centrality measure to identify the most influential spreaders~\cite{Erkol2019SystematicCB,Flavio2018}, since centrality can be defined in terms of distance, flow, and random walks~\cite{Erkol2019SystematicCB,Flavio2018}. Recently in \cite{arnaudon2019graph}, the authors introduce a scale-dependent multiscale centrality underlying the geometry of the network, by the reachable of nodes with the simple diffusion dynamic governed by the Laplacian of the graph. However, the centrality should be adapted to more complex dynamics, e.g., epidemic or rumor spreading. Moreover, there are not consolidated results or analyzes of this new method in terms of the IMP.

After defining the centrality measure, the influence maximization method selects the $\eta$ most central vertices as the initial spreaders (see Fig.~\ref{fig:schema}). Like in previous works~\cite{Vega-OliverosH17,Kitsak2010, Kandhway2016}, here we consider the well adopted degree (\emph{DG}), betweenness centrality (\emph{BE}) and PageRank (\emph{PR}) to measure the centrality of each node.

Another heuristical approach considered here is based on the network community structure. A community is a group of nodes that has more connections between them than with nodes in other groups. In social networks, communities represent people that share affinities, defining the phenomenon of homophily~\cite{Weng2013}. This condition is the reason that information or sentiments propagate better in a community, making people more open to the information shared by their peers~\cite{Weng2013,Vega-OliverosASONAM17}. Few works have considered the community structure in the influence maximization problem~\cite{Galstyan2009, Wang2010, Weng2013,Cao2011, Zhang2013, Vega2015spreader, Hosseini-Pozveh2016}.  Here, we propose a new approach to select the initial spreaders by the most central vertices within each community, i.e., we avoid selecting two nodes in the same community. 

As depicted in Fig.~\ref{fig:schema}, the main $\eta$ communities of the network are detected and for each isolated $G'_{i}$ community, we calculate a centrality measure and select the most central vertex within the community. Notice that we set the number of influential spreaders $\eta$ and the communities obtained may not correspond to the best division of the network into communities, which yields the maximum modularity. Therefore, we obtain $\eta$ seeds in which the influence overlap is minimized, but the influence within communities is maximized. Several methods have been developed to detect the communities on networks~\cite{Fortunato16,Wang2010}. Here we employ the {fastgreedy} algorithm, which is a fast and accurate method for community identification~\cite{Fortunato16, newman2004}. 

Similarly, in \cite{ghalmane2019centrality} the authors presented a measure considering two types of influence for the nodes in terms of the community structure: the local influence -- linked to the intra-community centrality, and the global influence -- related to the inter-community centrality. Their results, on top of {\SIR} simulations in a super-critical regime with $\lambda= 1.0$, confirmed that node rankings based on the community centrality are more accurate than standard/global centrality measures, in terms of the epidemic outbreak. However, part of the steps from the approach in \cite{ghalmane2019centrality} were originally presented here \cite{Vega2015spreader, Vega-Oliveros2017c}. Hence, their results are in accordance with our findings.

We also consider an approximation method based on a greedy hill-climbing algorithm~\cite{Kempe15}. Many approaches have been derived from this general greedy method~\cite{Kempe15}, such that most of them try to reduce the computational complexity to some polynomial order~\cite{Kempe15}. Here, we consider only the general greedy method~\cite{Kempe15}. The algorithm determines among all vertices $\textbf{s}_i \in V \backslash \Sd$, i.e. $\{\textbf{s}_i \in V \, | \, \textbf{s}_i \notin \Sd  \}$, the node that maximizes the function $\Inf(\Sd \cup \{\textbf{s}_i\})$, recalling that {\Sd} is initially empty. Afterwards, the vertex  $\textbf{s}_i$ is added to the set of seeds $\Sd = \Sd \cup \{\textbf{s}_i\}$ and the procedure runs until the target set achieves the size $|\Sd| = \eta$. 

\begin{table}[!tb]
	\centering
    \caption{Topological measures of the networks considered here.
   $\rho$ is the assortativity coefficient, $N$ the network size, $\langle k \rangle$ the average degree, $\langle g \rangle$ is the average shortest path length and $\langle C_c \rangle$ is the average clustering coefficient. The community-related parameters are the modularity $Q$ and the number of communities $N_c$.
   }
		\begin{tabular}{c|c|c|c|c|c|c c}
		\hline
Network & $\rho$ & $N$ & $\left\langle k \right\rangle$ & $\left\langle g \right\rangle$ & $\langle C_c \rangle$ & \multicolumn{2}{c}{FastGreedy} \\
 & & & & & & $Q$ & Nc\\  
 \hline
 \emph{BA} & $-0.43$ & $1000$ & $11.9$ & $2.96$ & $0.017$ & $0.26$  & $8$ \\
\emph{BA} & $-0.31$ & $1000$ & $11.9$ & $2.87$ & $0.028$ & $0.25$ & $9$ \\
\emph{BA} & $-0.21$ & $1000$ & $11.9$ & $2.86$ & $0.031$ & $0.25$ & $9$ \\
\hline
\emph{BA} &  $0.02$ & $1000$ & $11.9$ & $2.91$ & $0.035$ & $0.25$ & $10$ \\	
\emph{BA} &  $0.11$ & $1000$ & $11.9$ & $2.94$ & $0.034$ & $0.25$ & $11$ \\
\emph{BA} &  $0.34$ & $1000$ & $11.9$ & $3.11$ & $0.026$ & $0.27$ & $8$ \\
\hline
\emph{MSF} &  $-0.21$ & $1000$ & $11.6$ & $4.05$ & $0.390$ & $0.84$ & $14$ \\
\emph{MSF} &  $-0.13$ & $1000$ & $11.4$ & $4.36$ & $0.381$ & $0.86$ & $21$ \\
\emph{MSF} &  $-0.06$ & $1000$ & $12.8$ & $3.48$ & $0.502$ & $0.82$ & $10$ \\
\emph{MSF} &  $-0.15$ & $1000$ & $12.0$ & $3.23$ & $0.286$ & $0.59$ & $12$ \\
\hline 
\emph{Google+} & $-0.39$ & $23613$ & $3.32$ & $4.03$ & $0.174$ &  $0.74$ & $33$ \\
\emph{Internet} & $-0.20$ & $22963$ & $4.22$ & $3.84$ & $0.231$ &  $0.63$ & $57$   \\
\emph{Caida} & $-0.20$ & $26475$ & $4.03$ & $3.87$ & $0.208$ &  $0.64$ & $43$   \\
\emph{Advogato} & $-0.09$ & $5054$ & $15.6$ & $3.27$ & $0.253$ & $0.34$ & $49$ \\	
\hline
\emph{email} & $0.01$ & $1133$ & $9.62$ & $3.60$ & $0.220$ & $0.49$  & $16$ \\
\emph{Hamsterster} & $0.02$ & $2000$ & $16.1$ & $3.58$ & $0.539$ & $0.46$ & $57$ \\
\emph{PGP} & $0.23$ & $10680$ & $4.55$ & $7.48$ & $0.266$ & $0.85$ & $179$ \\
\emph{Astrophysics} & $0.23$ & $14845$ & $16.1$ & $4.79$ & $0.638$ & $0.63$ & $1172$ \\ \\
\hline
\end{tabular}
	\label{tab:basesUtilizadas}
\end{table}

\section{Databases}

We perform extensive numerical simulations in several artificial and real-world networks, evaluating the impact of the degree correlation in the influence maximization problem. The structural properties of the networks are summarized in Table~\ref{tab:basesUtilizadas}, with the respective assortativity $\rho$, number of vertices $N$, average degree $\left\langle k \right\rangle$, average shortest path length $\left\langle g \right\rangle$ and the average clustering coefficient  $\langle C_c \rangle$. Also, the highest modularity $Q$ value and number of communities $N_c$ identified by the fastgreedy algorithm are reported. We can see that real-world networks are naturally more modular than artificial networks (except for the modular scale-free artificial networks \emph{MSF}). However, most networks present similar values of the average shortest path length ($\left\langle g \right\rangle$). 

\subsection{Artificial networks}

We employ the algorithm proposed by \citeauthor{Xulvi-Brunet2005}~\cite{Xulvi-Brunet2005} for controlling the degree-degree correlation in Barab{\'a}si-Albert (BA) networks. This algorithm performs rewirings in order to increase or decrease the degree-degree correlation, i.e., by favoring the connection between nodes with similar degrees, or by hubs and low degree nodes. For controlling the modularity of the network, we employ the benchmark implementation for hierarchical and modular networks proposed by \citeauthor{Fortunato2009}~\cite{Fortunato2009}. This algorithm receives as inputs the number of nodes, average and degree distribution, the minimum and maximum number of communities, the level of mixing or overlapping among the communities. As results, we obtain scale-free networks with different levels of modularity, i.e., community structure. We will refer to this modular scale-free networks as ``MSF'' networks. We adopt these particular models as social networks also present scale-free degree distribution, community structure, and degree-degree correlation.

\subsection{Real world networks}

We consider eight real-world datasets representing connections in social and communication networks. The disassortative networks are (i) \emph{Google+}~\cite{GplusPaper}, which is an user-user social network; (ii) \emph{Internet}~\cite{NewmanDataSet}, which represents a fraction of the symmetrical snapshot of the Internet structure at the level of autonomous systems, reconstructed from BGP tables published on \textsc{routeviews.org} project; (iii) \emph{Caida}~\cite{Leskovec2005}, which is an undirected network whose nodes are autonomous system on the Internet, collected in 2007 from the CAIDA project; and (iv) \emph{advogato network}~\cite{massa09:advogato}, which is an online platform for free software community launched in 1999 that considers trust relationship between developers. The assortative networks are (i) the \emph{email network}~\cite{email:guimera}, which is a network of emails exchanged  between members of the \textit{Rovira i Virgili} University; (ii) \emph{hamsterster}~\cite{konect:hamster}, which is a network based on the friend and family relationship among users of the \textsc{hamsterster.com} website; (iii) \emph{PGP}~\cite{Boguna2004}, which is the largest component of the network of users of the Pretty-Good-Privacy algorithm for secure information interchange; and (iv) \emph{astrophysics}~\cite{astrophysics:newman}, which is a collaborative network between scientists on previous studies of astrophysics on arXiv. We assume that all these networks are undirected and unweighted. Only the largest network component is considered in our analysis.

\section{Results and discussion}

\subsection{Impact of assortativity on artificial networks}

\begin{figure*}[!bht]
\centering
\subfigure[MSF ($\rho=-0.21$, $Q=0.84$)]{\includegraphics[scale=0.23]{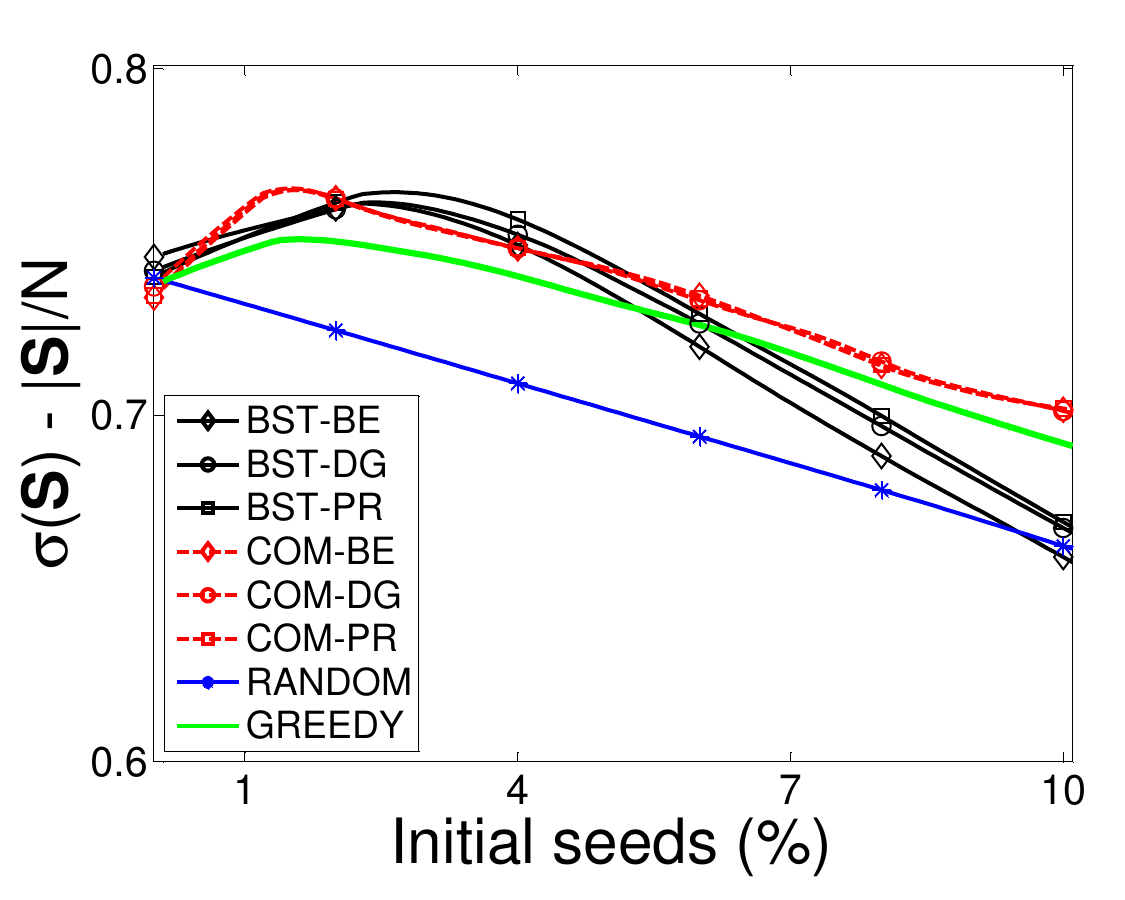}}  
\subfigure[MSF ($\rho=-0.13$, $Q=0.86$)]{\includegraphics[scale=0.23]{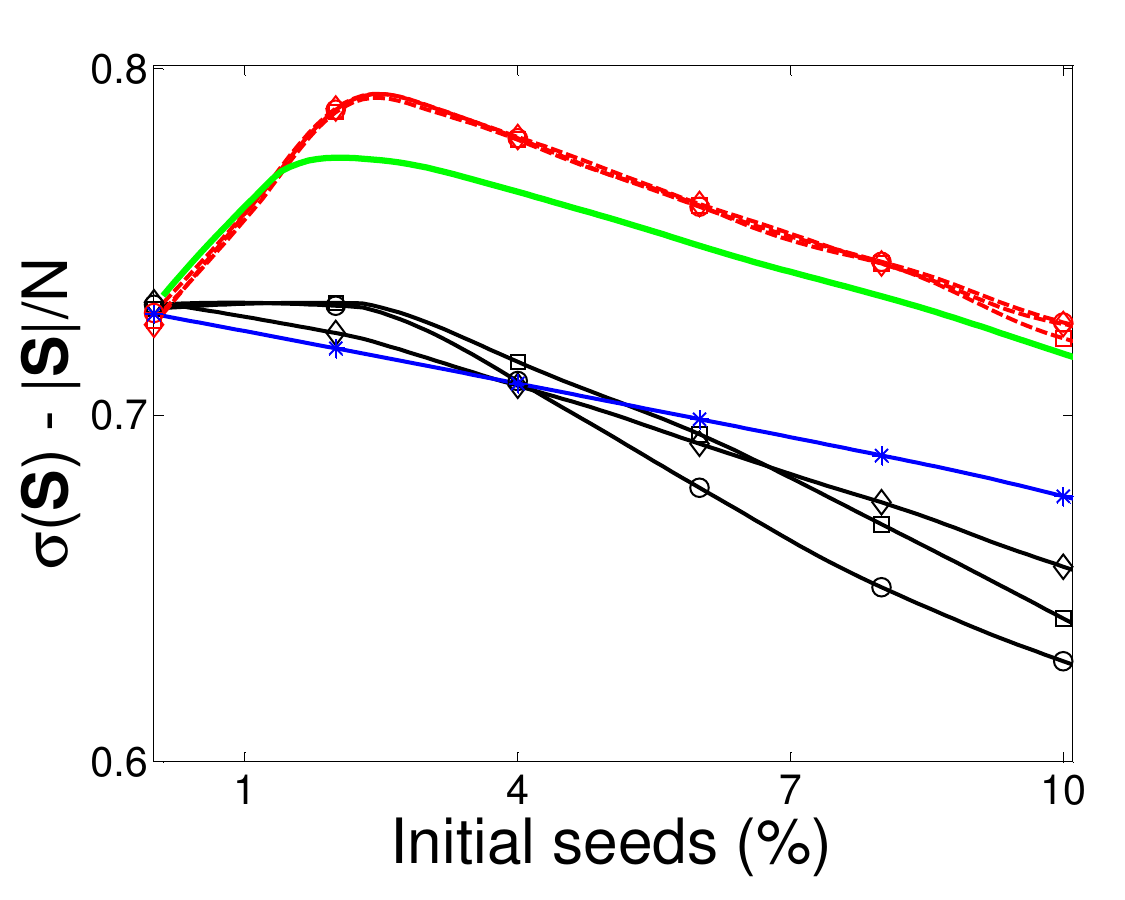}}  
\subfigure[MSF ($\rho=-0.06$, $Q=0.82$)]{\includegraphics[scale=0.23]{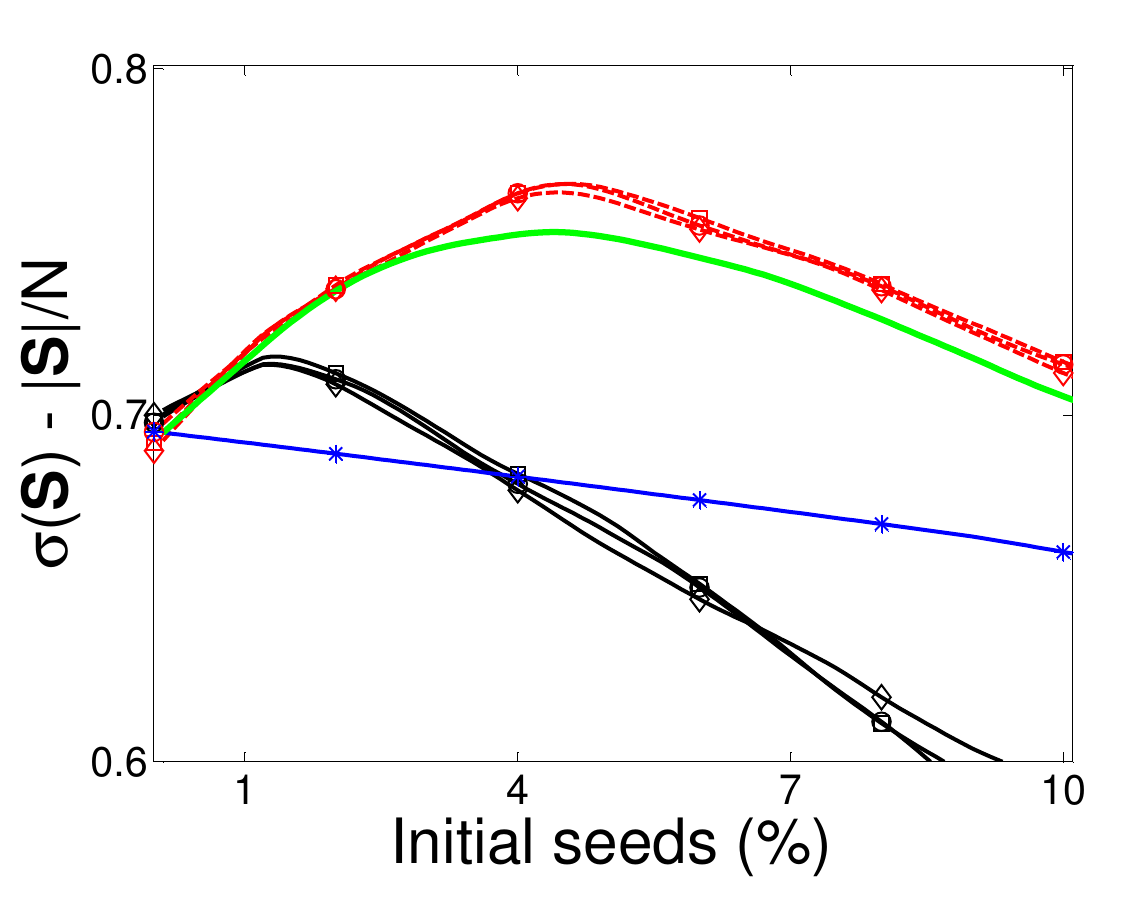}}   
\subfigure[MSF ($\rho=-0.15$, $Q=0.59$)]{\includegraphics[scale=0.23]{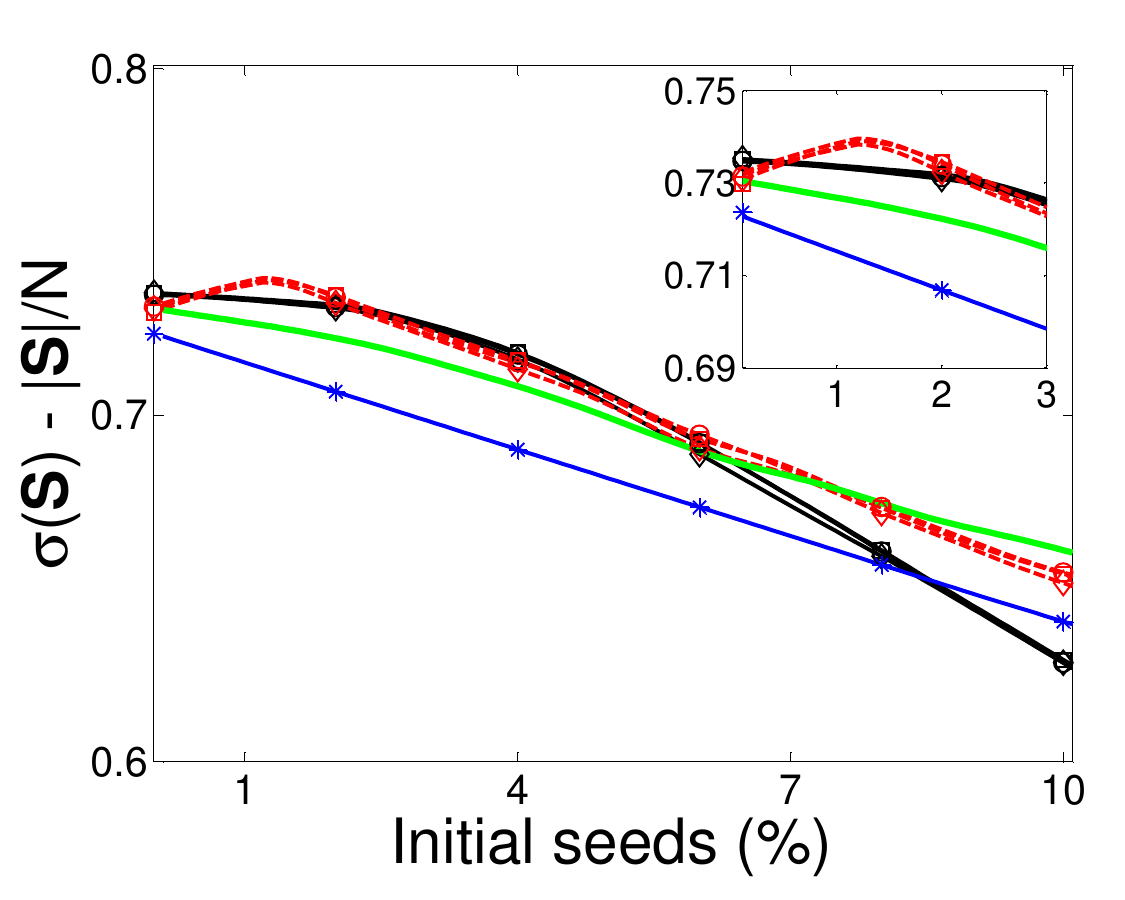}} 
\subfigure[BA ($\rho=-0.43$, $Q=0.26$)]{\includegraphics[scale=0.23]{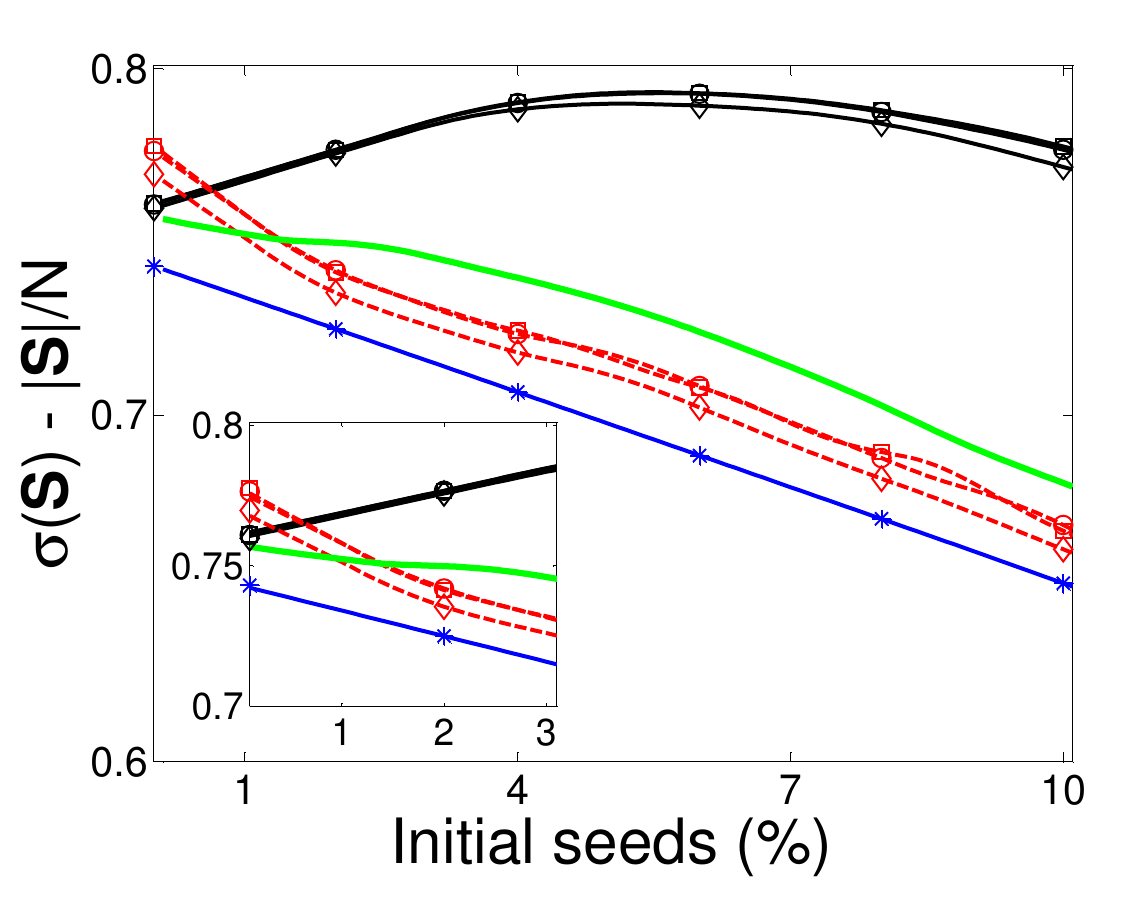}}  
\subfigure[BA ($\rho=-0.31$, $Q=0.25$)]{\includegraphics[scale=0.23]{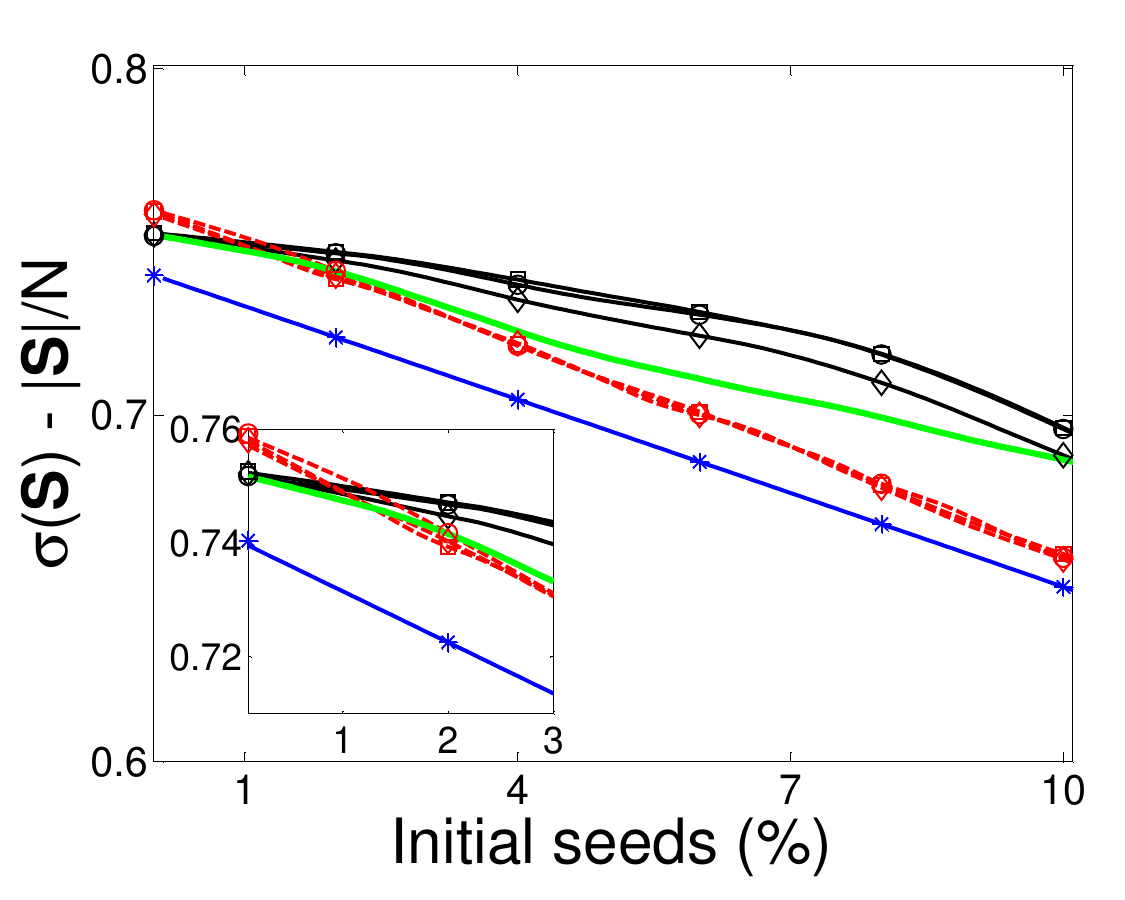}}  
\subfigure[BA ($\rho=0.02$, $Q=0.25$)]{\includegraphics[scale=0.23]{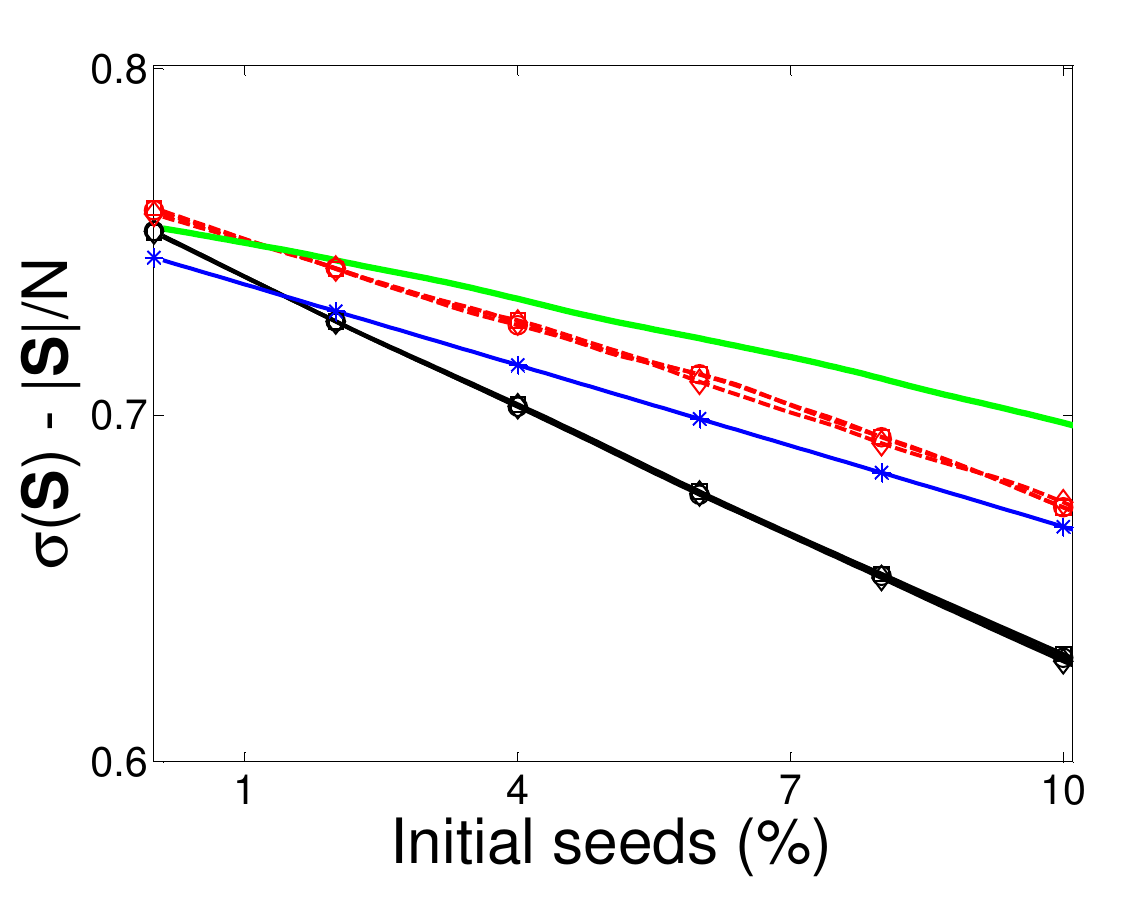}}  
\subfigure[BA ($\rho=0.11$, $Q=0.25$)]{\includegraphics[scale=0.23]{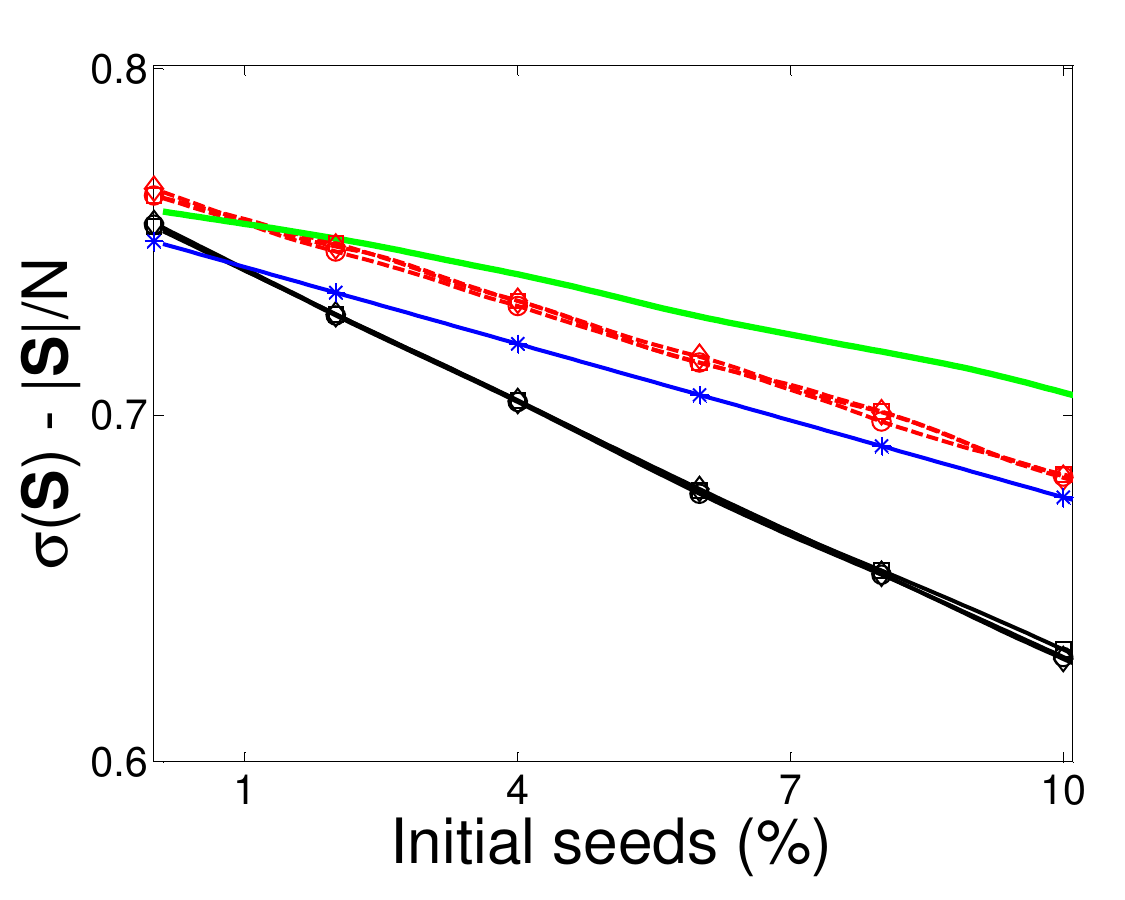}}
\caption{Impact of degree-degree correlation on the influence maximization problem, when maximizing the relative size of the outbreak in function of the initial seeds (\Sd). The number of seeds varies from two nodes to $10\%N$. For each artificial scale-free network, we calculate the set of initial seeds according to: (BST) the best-ranked nodes of the network; (COM) the most central nodes from communities; (RANDOM) randomly selecting the initial seeds; and (\emph{GREEDY}) the greedy method. The adopted measures are betweenness centrality (\emph{BE}), degree (\emph{DG}) and PageRank (\emph{PR}). The inset figures are the zoom at low number of seeds $|\Sd| = 3\%N$, with the same axes units. }
\label{fig:resArtf}
\end{figure*}

We calculate the final fraction of influenced individuals according to Equation~\ref{eq:infFunc}. The number of initial seeds varies from two nodes to $10\%$ of the total number of nodes ($N$). The impact of degree correlation on the influence maximization problem is illustrated in Fig.~\ref{fig:resArtf}, where we show the relative maximum spreading, i.e., the fraction of maximum informed nodes within the population that was not initially informed (initial seeds). We observe that an unexpected phenomenon occurs when networks are disassortative (see Fig.~\ref{fig:resArtf} with negative $\rho$) --- the curve of relative maximum influenced nodes ($\Inf(\Sd) - |\Sd|/N$) has a peak when the number of seeds is lower than $10\%$ of the network and then starts to decline when the number of seeds is increased. This peak is due to the low interaction between the initial spreaders in disassortative networks, i.e., central nodes (e.g. hubs) are connected through low degree nodes, increasing the distance between them or branching~\cite{EstradaAssorta2011}. If the number of seeds is higher than $10\%$ of the total number of nodes, then the overlap of influence between spreaders occurs and they become stiflers more frequently than for a smaller number of seeds. 

We also observe that selecting the central nodes by the communities provides better results than the greedy method when the networks have high modularity (see Fig.~\ref{fig:resArtf}(a)-(d)). For instance, in the disassortative networks, the community detection approach works well for high modularity values. This is since communities are well defined and the centroids are well separated, which can be observed by the high modularity and disassortative property. Therefore, there is a lower effect of spreading overlapping between the community seeds.

Another interesting point is the crossover phenomena at low numbers of initial seeds, as shown in the inset of Fig.~\ref{fig:resArtf}(d)-(f). We observe that there exist a maximum relative size of the outbreak for the community approach, which happens in networks with a well defined modular structure. When the modularity is too small, a small number of initial seeds achieved larger relative outbreak size, which decreases with the size of $\Sd$. On the other hand, the higher ranking strategy improves the relative maximum outbreak size when additional seeds are considered, and the reach of the outbreak peak is enhanced in stronger disassortative networks.

For non-correlated and assortative networks (please, see Section~\ref{sec:Method} for network correlation definitions), selecting the central nodes by considering the whole network leads to the lowest relative size of the outbreak ($\Inf(\Sd) - |\Sd|/N$) and the results are worse than the uniform selection of seeds (see the curves of random in Fig.~\ref{fig:resArtf}(c), (g), (h) and Fig.~\ref{fig:resArtfAnnex}(b)). This happens because hubs are placed in the same community and, therefore, there is a large number of interactions between the initial spreaders, reducing the rumor propagation.
The artificial BA networks considered are less modular than real-world networks (see Table~\ref{tab:basesUtilizadas}), which makes the algorithm by communities performing worse than the greedy seed selection.

Additionally, in  Fig.~\ref{fig:resArtf}, we observe that the selection according to communities yields similar results for all centrality measures considered. The same happens to the selection according to global centrality (best-ranked approach). Thus, the selection among the centrality measures does not affect the prediction of the fraction of influenced nodes significantly. This can be explained by the high correlation between degree, PageRank and betweenness centrality in scale-free networks, which have been reported as producing statistical similar results in other domains~\cite{Schoch_2017,Vega-Oliveros2019b}. However, one could consider more elaborated centrality measures that outperform heuristic methods in the identification of influential nodes~\cite{Vega-Oliveros2019b, Chen2019, Tixier2019arxiv, LU20161, ZHOU201969,arnaudon2019graph}. For instance, combining local topology information into a multi-centrality index~\cite{Vega-Oliveros2019b}; synthesizing local features and spreading influence in a fusion~\cite{Chen2019} or multiscale~\cite{arnaudon2019graph} index; or combining a node centrality from perturbed versions of the network~\cite{Tixier2019arxiv}.

Previous results suggest that the reach of the rumor depends on the network assortativity and community organization, the method for selecting the initial spreaders, and the number of seeds. For instance, in \ref{sec:appendixA} we have different results among the methods in two networks with similar structural properties, but opposite degree-degree correlation. Another example is the greedy approach in Fig.~\ref{fig:resArtf}(e), which could not follow the growing behavior of the higher ranking strategy. This dependency is not expected, since the number of infected nodes should always increase with the number of seeds. The before is approximately observed in the linear behavior for the uniform selection of initial spreaders, without considering the network topology.

\subsection{Impact of assortativity on real world networks}

We also analyze the spreading process in eight real-world networks (see Table~\ref{tab:basesUtilizadas}). These networks present different levels of degree-degree correlation and community organization. The optimal fraction of stiflers $\Inf(\Sd)$, i.e., the maximum informed individuals in function of {\Sd}, is calculated by considering the number of initial seeds in the interval $[2, \eta_{\hbox{\small max}}]$, where $\eta_{\hbox{\small max}}$ corresponds to  $10\%$ of nodes. We limit the simulation until $10\%$ of network size because the greedy method has a very high computational cost. For instance, the identification until the $10\%$ of most influential spreaders took about two months for the \emph{Caida} network, whereas our method provides similar results in less than two hours.

\begin{figure*}[!btp]
\centering
\subfigure[Google+ ($\rho=-0.39$)]{\label{fig:googleTP}\includegraphics[scale=0.23]{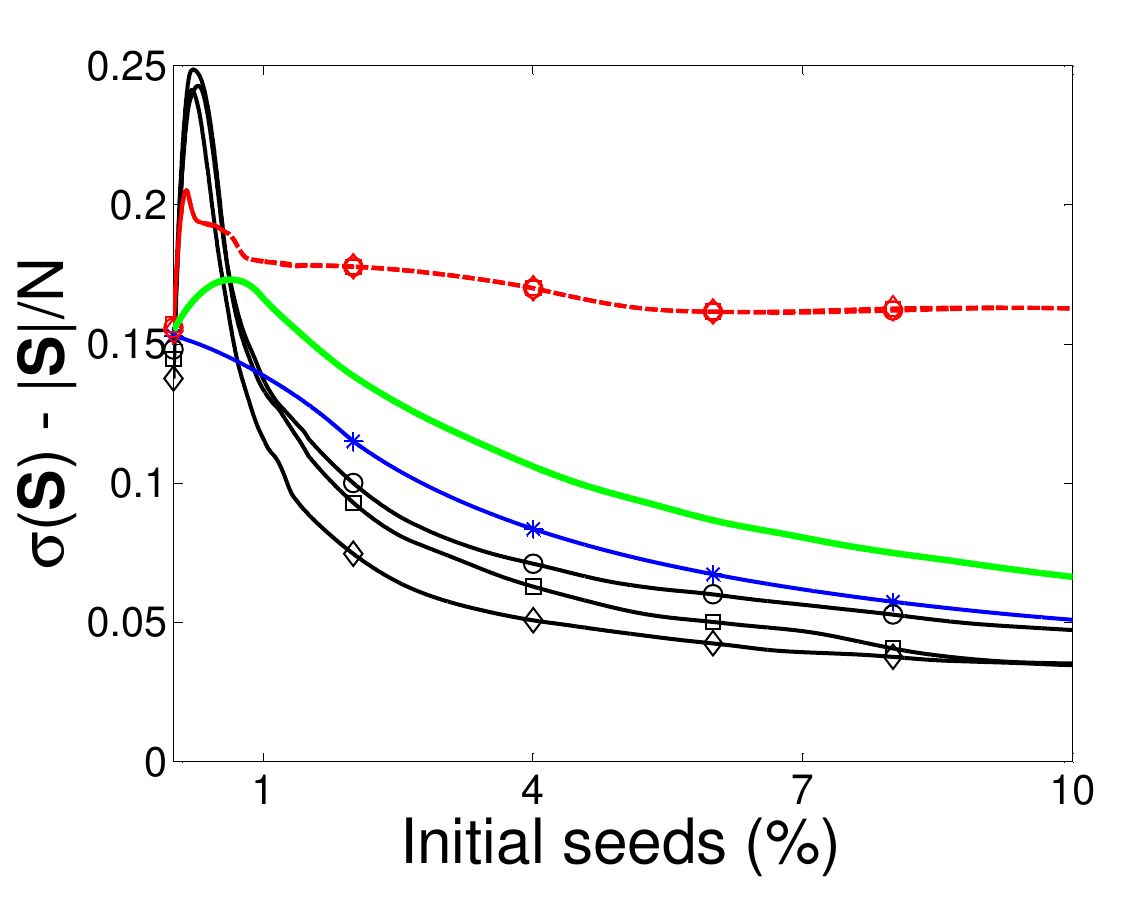}}
\subfigure[Internet ($\rho=-0.20$)]{\label{fig:internetTP}\includegraphics[scale=0.23]{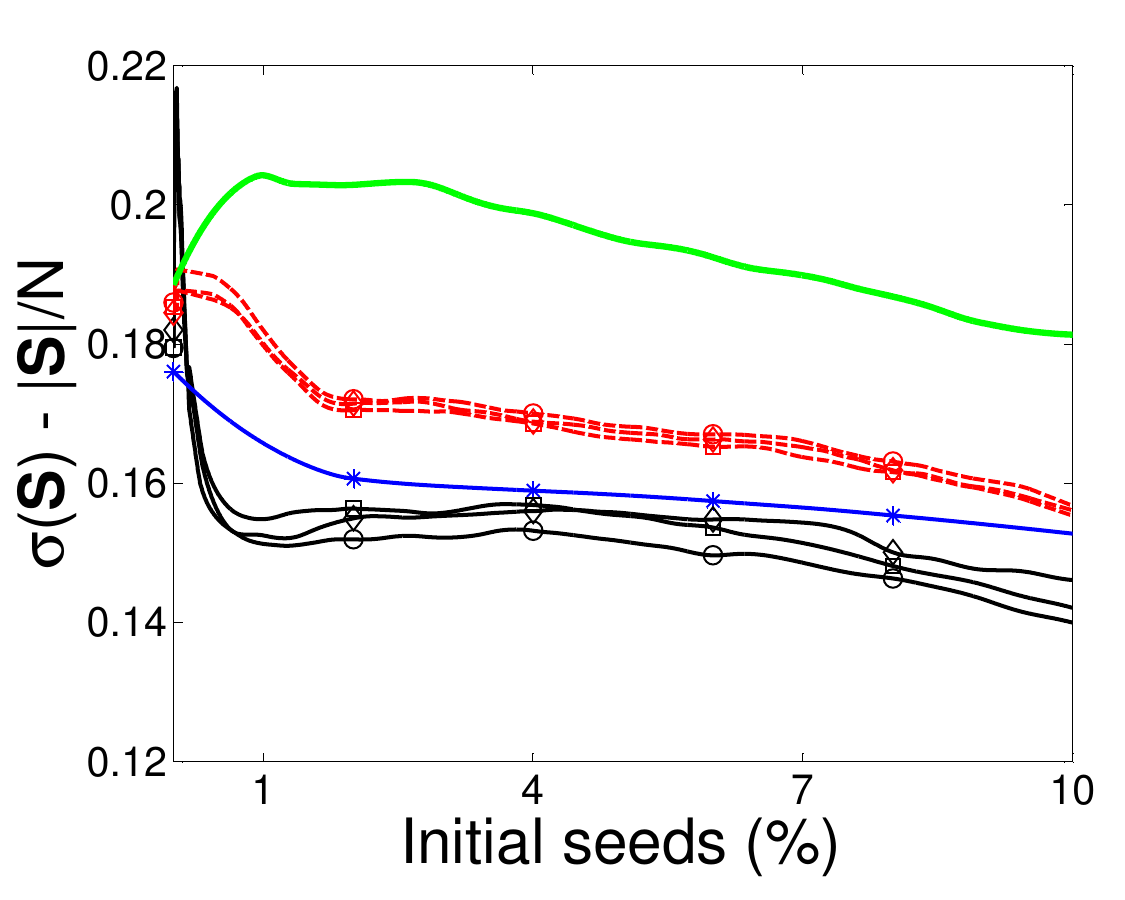}} 
\subfigure[Caida ($\rho=-0.20$)]{\label{fig:caidaTP}\includegraphics[scale=0.23]{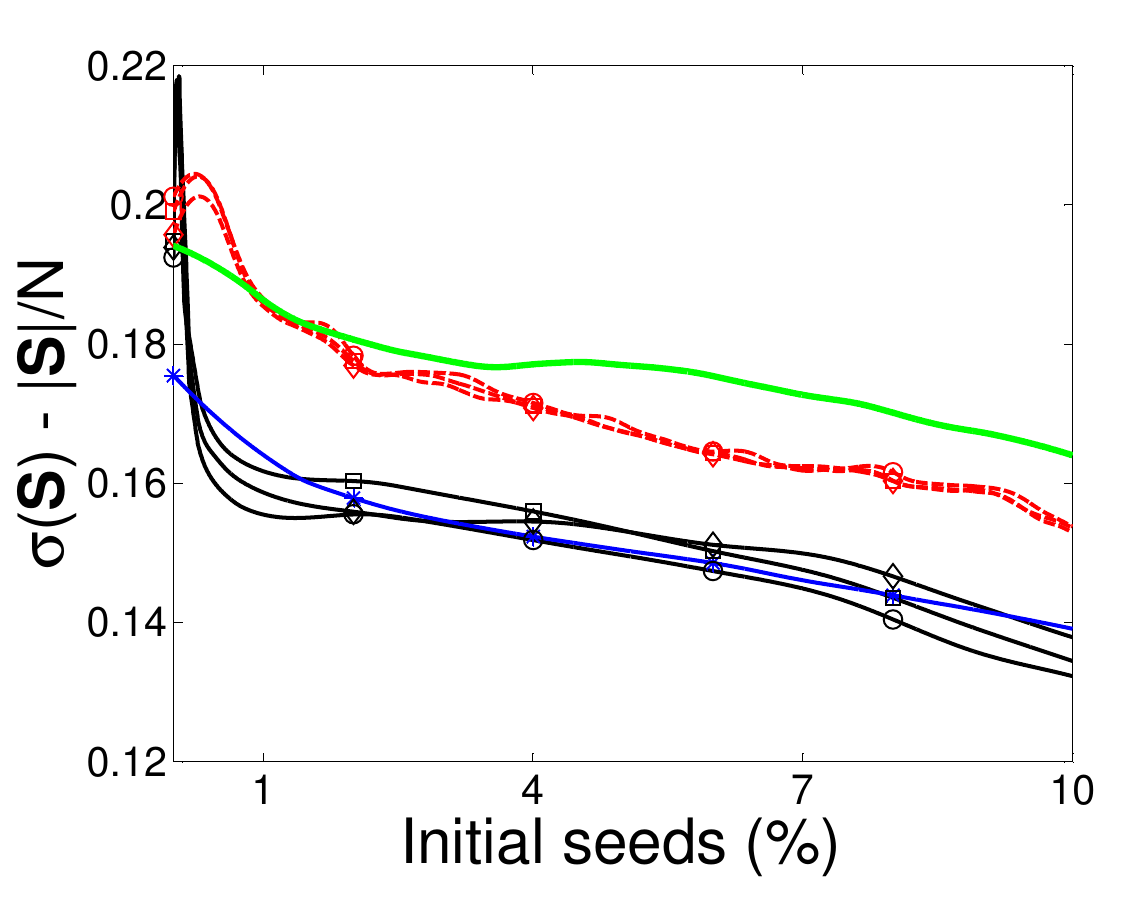}} 
\subfigure[Advogato ($\rho=-0.09$)]{\label{fig:advogatoTP}\includegraphics[scale=0.23]{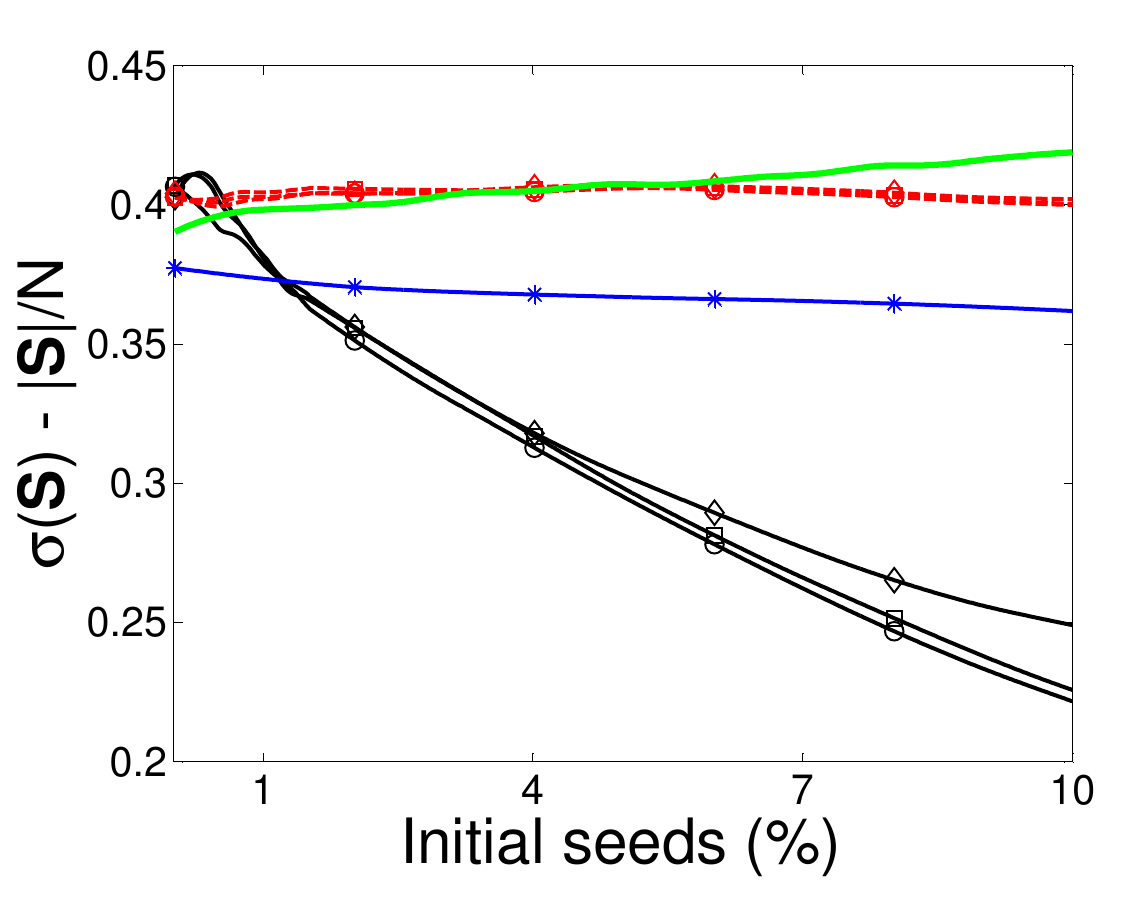}}
\subfigure[email ($\rho=0.01$)]{\label{fig:emailTP}\includegraphics[scale=0.23]{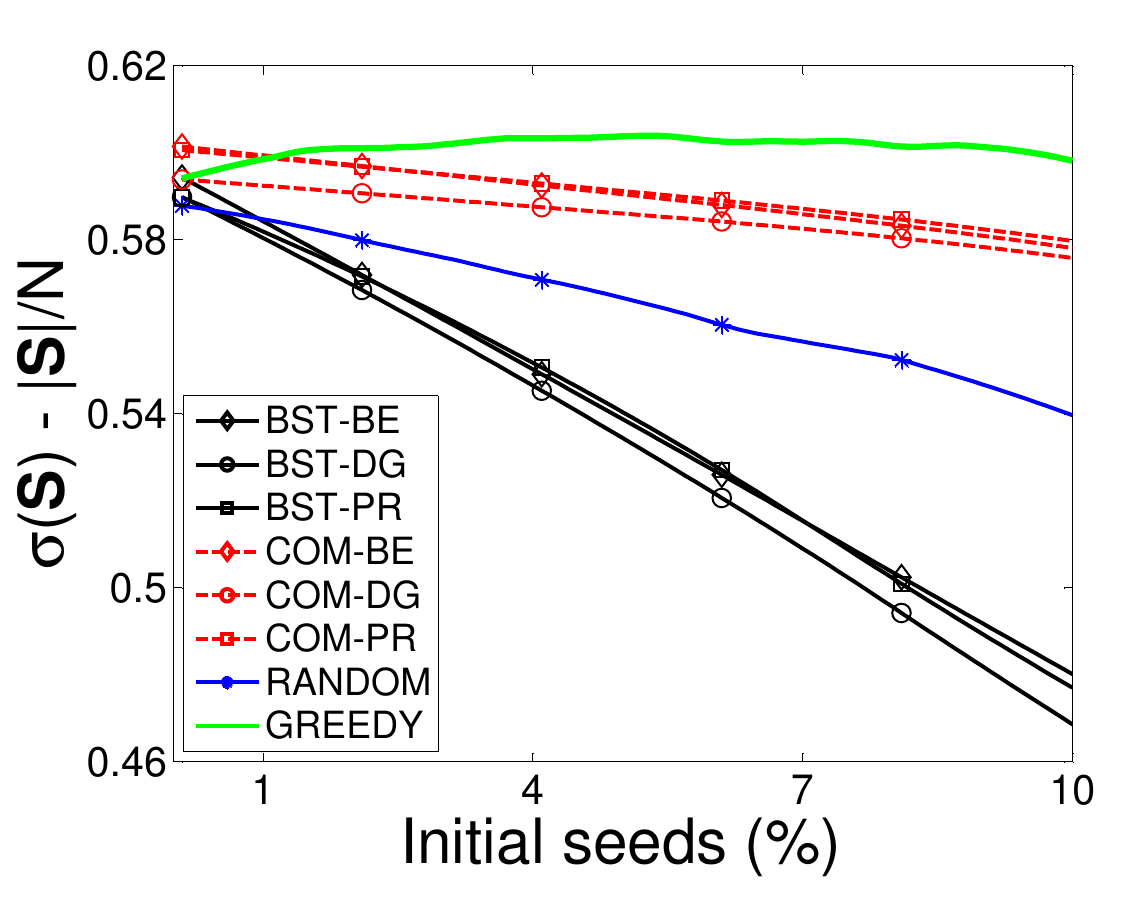}}    
\subfigure[Hamsterster ($\rho=0.02$)]{\label{fig:hamstersterTP}\includegraphics[scale=0.23]{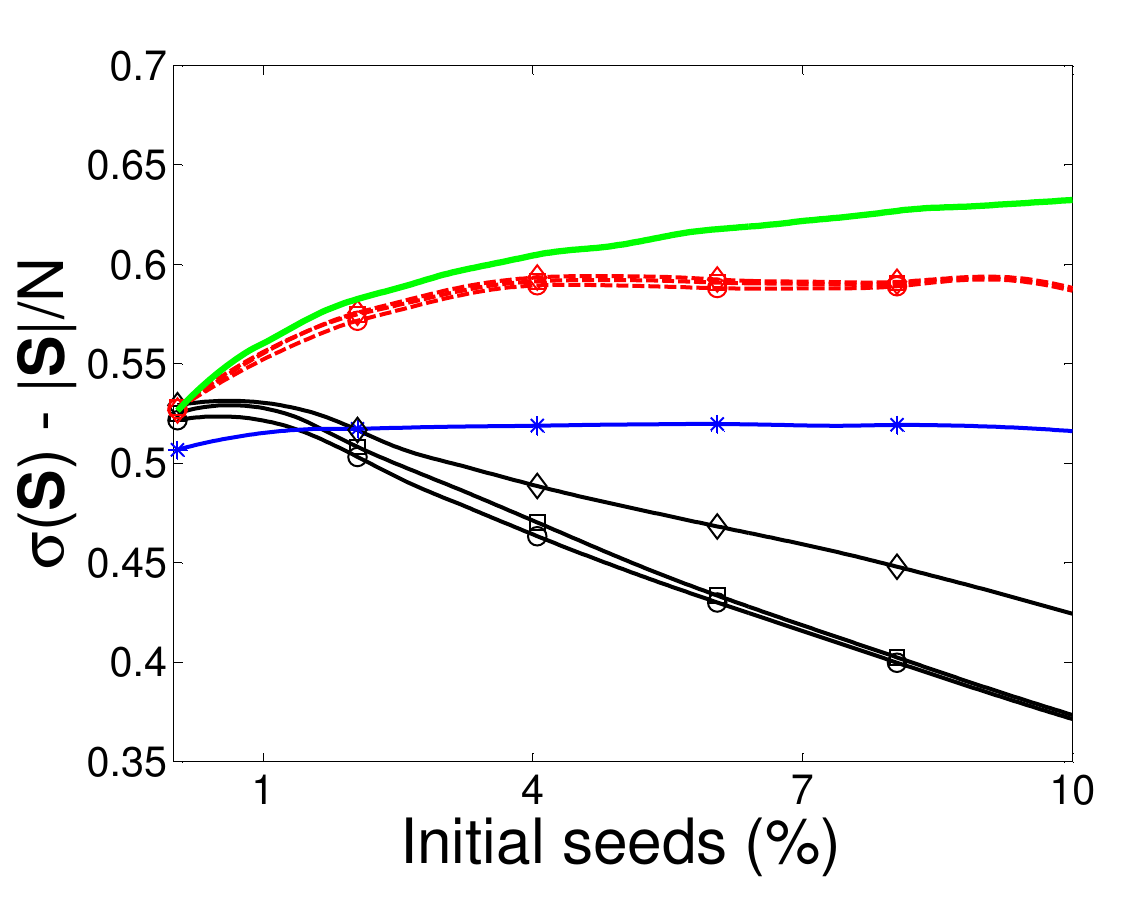}}  
\subfigure[PGP ($\rho=0.23$)]{\label{fig:PGPTP}\includegraphics[scale=0.23]{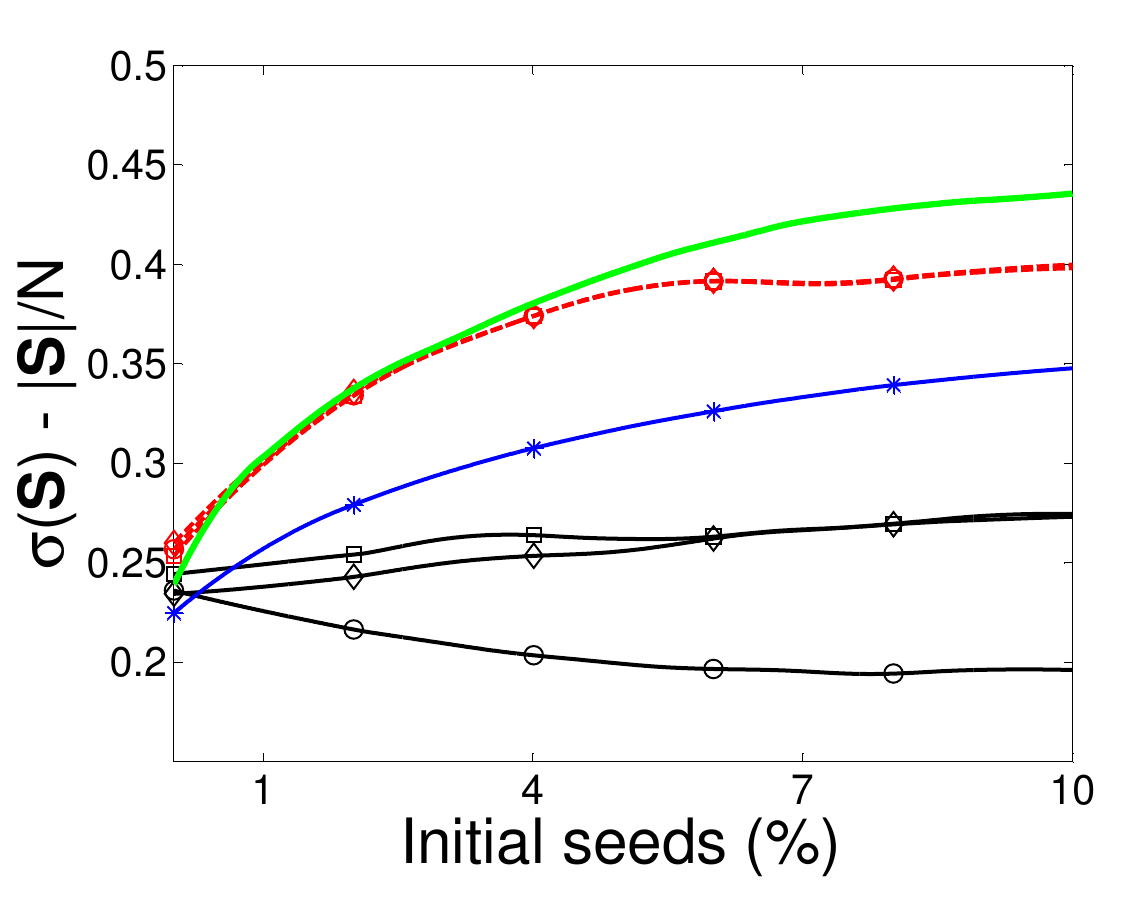}}
\subfigure[Astrophysics ($\rho=0.23$)]{\label{fig:astrophysicsTP}\includegraphics[scale=0.23]{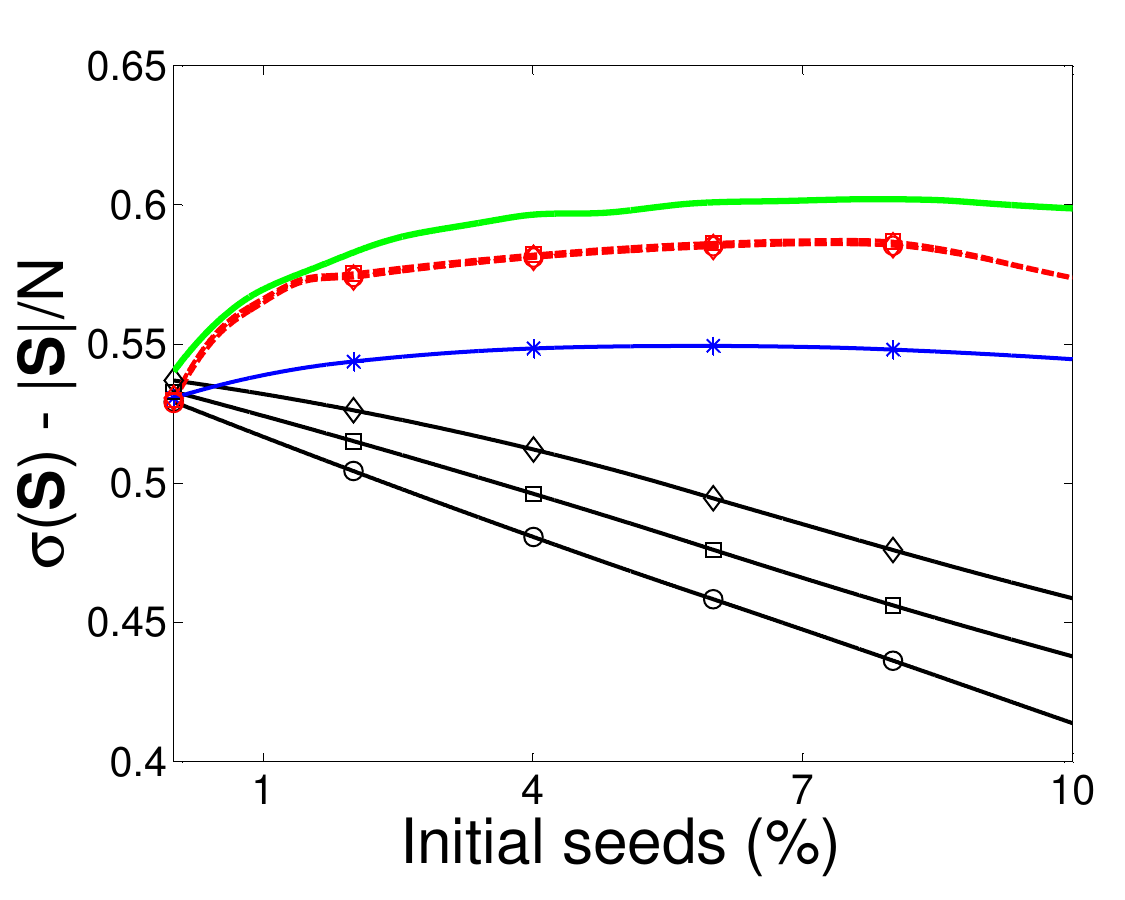}}
\caption{\label{fig:resReals} Impact of degree correlation in the real-world networks, when maximizing the relative size of the outbreak in function of the initial seeds (\Sd). The number of seeds varies from two nodes to 10\% of network size, selected according to communities (\emph{COM}), ranking of the most central nodes (best-ranked, \emph{BST}), random (\emph{RANDOM}) and by the greedy method (\emph{GREEDY}). Betweenness centrality (\emph{BE}), degree (\emph{DG}) and PageRank (\emph{PR}) are the centrality measures used.}
\end{figure*}

The set of initial spreaders are selected uniformly at random or according to centrality measures calculated from the whole network or inside communities. Fig.~\ref{fig:resReals} shows the maximum relative fraction of stiflers obtained in function of the number of seeds. We can see that a peak in the relative curve ($\Inf(\Sd) - |\Sd|/N$) occurs in disassortative networks for the higher ranking strategy. These results are similar to those obtained in artificial networks (see Fig.~\ref{fig:resArtf}).

Contrary what is currently expected, the choice of seeds according to global centrality measures provides lesser informed nodes than the random selection of seeds, mainly in assortative networks (see for instance Fig.~\ref{fig:resReals} (e),(g),(h) and Fig.~\ref{fig:resArtfAnnex}(b)). This is due to the interaction of central spreaders in the early steps of the process, as central nodes tend to be connected in these networks and cause that spreaders in rumor dynamic turn inactive more quickly.

Fig.~\ref{fig:resReals} also shows that the selection of initial spreaders according to communities provides a larger fraction of informed individuals than the greedy approach for the Google+, and at small sizes of initial seeds for Caida network. Generally, the greedy method tend to have poor performance for low sizes of {\Sd} when more disassortative and modular the networks. The before can be observed in Fig.~\ref{fig:resArtf} (a)-(e) and Fig.~\ref{fig:resReals} (a),(c)-(e).

For the remaining networks, the greedy approach is better, in most cases, than all the other methods. However, the relative maximum size of outbreaks is similar to the community-based results. Thus, the greedy strategy for real-world networks is harder concerning artificial models. This is due to the community structure present in real networks, with possible unbalance communities and intersections. This property could be a limitation for the fastgreedy community detection algorithm, which finds a well-separated dendrogram of the communities. However, since the greedy method is computationally expensive, the selection of seeds according to communities reveal to be more suitable. 

\subsection{Distribution of initial spreaders}

We analyzed how the distribution of the seeds is similar in terms of communities. Since these communities have different sizes, we aim to calculate the probability of uniformly select a vertex that belongs to some $G'_{c}$ community, i.e.,  $p(\mathbi{x}_c) = |G'_{c}|/N$, where $|G'_{c}|$ is the number of vertices in the community. 
In this way, we define $\setX = \{ \mathbi{x}_{1}, \mathbi{x}_{2}, \ldots, \mathbi{x}_{\eta} \}$ as the set that describes the probabilities $p(\mathbi{x})$ for all the communities. Moreover, we say that $\mathbi{y}_{c} = \Sd \cap G'_{c}$ denotes the seeds that are part of a community. Thus, the probability of uniform select a seed that belongs to some community $G'_{c}$ is $p(\mathbi{y}_c) = |\mathbi{y}_c|/\eta$. In the same way, $\setY = \{\mathbi{y}_{1}, \mathbi{y}_{2}, \ldots, \mathbi{y}_{\eta}\}$ is the set that describes the probabilities $p(\mathbi{y})$ for all the communities.

We expect that rumor spread better when the seeds are distributed among communities. This should occur because each seed tries to infect its own community and the interaction between pairs of spreaders is minimized. We verify this hypothesis here by the normalized variation of information (\NVI), inspecting the similarity between the size of communities and the distribution of the seeds in the communities. The {\NVI} is an information-theoretic metric that obeys the triangle inequality~\cite{Vinh2010} and is normalized in a stochastic sense defined as follow:
Given two discrete sets $\setX$ and $\setY$, their joint information entropy ($\mathcal{H}$) and  mutual information ($\mathcal{I}$) are expressed respectively in terms of the marginal and joint distributions of  $\setX$ and $\setY$,
\begin{equation}
\NVI(\setX, \setY) = 1 - \frac{\mathcal{I}(\setX, \setY)}{\mathcal{H}(\setX, \setY)} ,
\label{eq:I}
\end{equation} 
where the results are normalized in $[0,1]$ and the mutual information measures the overlap between the two sets.  In information theory this measure returns $0$ when the two sets are identical in the information of the item distribution and $1$ when they are completely different, i.e., they are independent with no sharing of information.

In terms of the seed distribution across communities, we identify $\eta$ main communities and calculate the {\NVI} measure for the seeds according to the greedy approach and degree centrality, as shown in Table~\ref{tab:coeffV}. When the set of seeds is similarly distributed across communities, the {\NVI} values tend to zero. When the seeds are concentrated in only one partition, the {\NVI} values tend to a maximum equal to unity. Lower {\NVI} values mean that the seeds are more homogeneously distributed as the size of the communities.

\begin{table}[!bt]
	\centering
	\caption{The normalized variation of information ({\NVI}) measure calculated for the selection of seeds according to the greedy (GREEDY) approach or degree centrality (BST-DG).}
		\begin{tabular}{ c c r c c c }		
		\hline \hline
 	& Network & $\rho$ \phantom{a}  & $|\Sd|/N$ & { GREEDY }  & { BST-DG} \\ 
 	&       &  & (\%)    & {\NVI}  & {\NVI} \\  \hline
 	& \emph{BA} & $-0.43$ & $0.5$ & $0.409$ & $0.409$\\
  & \emph{BA}  & $-0.43$ & $1$   & $0.556$ & $0.535$\\
\DN & \emph{BA} & $-0.43$ & $10$ &$0.706$ & $0.731$\\
 	& \emph{google+} & $-0.39$ & $0.5$ & $0.311$ & $0.408$ \\
  & \emph{google+} & $-0.39$ & $1$   & $0.196$ & $0.334$ \\
 	& \emph{google+} & $-0.39$ & $10$  & $0.464$ & $0.725$ \\
 	\hline 	
 	& \emph{BA} & $0.02$ & $0.5$ & $0.409$ & $0.409$ \\
  & \emph{BA} & $0.02$  & $1$   & $0.469$ & $0.651$ \\
\NC	& \emph{BA} & $0.02$  & $10$  & $0.708$ & $0.843$ \\
	& \emph{email} & $0.01$ & $0.5$ & $0.368$ & $0.306$ \\
  & \emph{email} & $0.01$ & $1$   & $0.524$ & $0.695$ \\
 	& \emph{email} & $0.01$ & $10$  & $0.603$ & $0.809$ \\
 	\hline 	
 	& \emph{BA} & $0.34$  & $0.5$ & $0.344$ & $0.689$ \\
  & \emph{BA} & $0.34$ & $1$   & $0.492$ & $0.858$ \\
\AN & \emph{BA} & $0.34$ & $10$ & $0.693$ & $0.861$ \\ 
 	& \emph{astrophysics}  & $0.23$ & $0.5$ & $0.523$ & $0.940$ \\
  & \emph{astrophysics} & $0.23$ & $1$   & $0.589$ & $0.948$ \\
 	& \emph{astrophysics} & $0.23$ & $10$  & $0.558$ & $0.933$ \\	
	\hline
	\hline	
	\end{tabular}	
	\normalsize
	\label{tab:coeffV}
\end{table}

We notice that the {\NVI} values for the greedy algorithm are lower than the higher ranking strategy according to the degree. This suggests that the greedy approach tends to select the seeds more homogeneously distributed across communities -- most of the cases in seed sets with sizes lower than 1\% of the network.  In the artificial networks, the {\NVI} values of the degree approach increase with $|\Sd|$, because seeds tend to be placed in specific communities. Moreover, the {\NVI} differences between Greedy and BST-DG algorithms increase with assortativity. The reason is that in high degree-degree correlated BA networks, hubs are all placed in the same community, and they tend to be connected.

Similarly, when increasing the assortativity, the seeds identified by the greedy method are more homogeneously distributed across the communities than the \emph{BST-DG} seeds, which is more notably observed in the $\NVI$ values for the real-world networks. This result supports the hypothesis that the seeds of the greedy method are distributed according to the communities of the network, avoiding spreading overlapping between the seeds as a possible strategy~\cite{BalkanskiIS17}. This also indicates that the influence maximization problem in networks with community organization can be addressed by the identification of the most central nodes inside communities, instead of considering the greedy approach, which is computationally more expensive. In networks without community structure, the greedy algorithm, or the selection according to the most central nodes (globally) provides the best results, as observed in Fig.~\ref{fig:resArtf}.

However, notice that our results are valid only for the Maki-Thompson rumor model. In the case of {\SIR}, cascade or threshold models~\cite{Erkol2019SystematicCB,Pei2019InfluencerII}, the results may be different. The investigation of the effect of community structure on these models are an interesting topic for further research.

\subsubsection{Statistical analysis}

Since the greedy approach selects seeds uniformly among communities, it is expected that the community-based and greedy methods provide a similar number of informed nodes. Thus, we perform a statistical test to compare the performance of the four methods considered here — the final fraction of recovered individuals measures this performance. Initially, we analyze the influence maximization considering $1\%$ of the network as initial spreaders (see Table~\ref{tab:influence}). We consider a statistically significance test employing the Friedman and Nemenyi approach~\cite{Demsar2006}. The Friedman test is a non-parametric counterpart of the well-known ANOVA (analysis of variance), with the corresponding Nemenyi post-hoc test for comparing the average ranks of the algorithms. If the Friedman test rejects the null hypothesis of similar performance, we proceed with the Nemenyi post-hoc test for pairwise comparisons, verifying whether the differences in rank values are statistically significant. 

\begin{table}[!bt]
	\centering
	\caption{Final fraction of informed nodes (\Inf(\Sd)) according to different methods.}
		\begin{tabular}{ c c c c c c c }		
		\hline \hline
 	Network & \multicolumn{2}{c}{GREEDY} & \multicolumn{2}{c}{COM-DG} & BST-DG & RANDOM \\ 
 	& \Inf($1\%$) & \Inf($10\%$)  & \Inf($1\%$) & \Inf($10\%$) & \Inf($1\%$) & \Inf($1\%$) \\  \hline
\emph{Google+} & $0.1686$ & $0.1663$  & $0.1897$ & $0.2627$ & $0.1549$ & $0.1501$ \\
\emph{internet} & $0.2130$ & $0.2810$ & $0.1906$ & $0.2570$ & $0.1639$ & $0.1755$ \\
\emph{caida} & $0.1959$ & $0.2640$ & $0.1966$ & $0.2531$ & $0.1696$ & $0.1743$\\
\emph{advogato} & $0.4071$ & $0.5179$ & $0.4138$ & $0.5002$ & $0.3933$ & $0.3821$\\
\emph{email} & $0.6086$ & $0.6976$ &$0.6089$  & $0.6707$ & $0.5942$ & $0.5941$\\
\emph{hamsterster} & $0.5738$ & $0.7318$ & $0.5693$ & $0.6871$ & $0.5408$ & $0.5236$\\
\emph{PGP} & $0.3126$ & $0.5356$ & $0.3122$ & $0.4985$ & $0.2593$ & $0.2654$\\
\emph{astrophysics} & $0.5785$ & $0.6979$ & $0.5735$ & $0.6745$ & $0.5418$ & $0.5456$\\	
	\hline
	\hline	
	\end{tabular}	
	\normalsize
	\label{tab:influence}
\end{table}

The critical diagram representation suggested by Dem\v{s}ar~\cite{Demsar2006} provides a visual method to compare the results. In the diagram, a horizontal line represents the axis with the average rank values of the methods. In this axis, the lowest     (highest) ranked methods are on the left (right) side. Algorithms that are not significantly different from each other are connected through a bold horizontal line. The performance between methods is significantly different if their corresponding average ranks differ by at least the critical difference CD. The value of CD given by the Nemenyi test is presented at the top of the diagram.

According to the result of $\Inf(1\%)$ in the Table~\ref{tab:influence}, the chi-square statistics for the methods is $19.20$, and the critical value of the chi-square statistics with $3$ degrees of freedom at $95$ percentile is $7.81$. Thus, for the Freidman test using the chi-square statistics, the null-hypothesis that all methods behave similarly should be rejected. Moreover, we calculate the F-statistics of the methods, obtaining the value of $28.00$. With $3$ and $21$ degrees of freedom and at $95$ percentile, the critical value of the F-statistics is $3.07$, indicating that the null hypothesis should be rejected again. Therefore, the method does not provide statistically similar results. 

\begin{figure}[hb!t]
\centering
\includegraphics[scale=0.75]{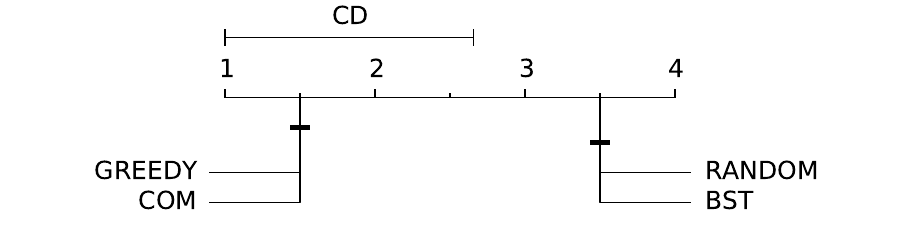}
\caption{\label{fig:nemenyi} The critical difference (CD), according to the Nemenyi test, for comparing the mean-ranking of two different methods at $95$ percentile is $1.66$. Mean-ranking differences above this value are significant and unconnected. }
\end{figure}

Since the methods do not provide the same fraction of informed nodes, we apply the post-hoc Nemenyi test in order to find which method achieves the maximum influence. The critical diagram of the Nemenyi test is shown in Fig.~\ref{fig:nemenyi}. The CD for comparing the mean ranking between two methods at $95$ percentile is $1.66$. Mean-ranking differences above this value are statistically significant. Thus, we conclude that there are no statistically significant differences in the influence maximization results between the greedy and community method when the number of seeds represents less than $1\%$ of the network. However, the Nemenyi test indicates significant differences between the methods based on community centrality and those that consider the random selection of spreaders or selection according to higher centrality.

We verify in Fig.~\ref{fig:resReals} that for $|\Sd| > 1\%N$, the methods based on the greedy method and community centrality provide the highest number of informed nodes. Thus, we perform the statistical hypothesis test only on the methods based on the greedy approach and community organization. For evaluation of these two algorithms in multiple data sets, we employ the Wilcoxon signed-rank test~\cite{Demsar2006}. This statistical test is a non-parametric alternative to the paired t-test. We adopt the Wilcoxon test because it is less affected to outliers and does not assume a particular population distribution~\cite{Demsar2006}. For rejecting the null-hypothesis of similar performance, the $W$-value returned by the test should be smaller than the corresponding critical $W_c$ value of the Wilcoxon test table. 

We compute the Wilcoxon statistical test at $95$ percentile for the greedy and community approach considering the number of initial spreaders as $|\Sd| = 10\%N$ (Table~\ref{tab:influence}). As a result, we obtain a $W$-value $= 8$. The critical value for eight networks at $p = 0.05$ is $W_c = 3$. Therefore, the null-hypothesis of similar performance of the methods cannot be rejected. These results suggest that the fraction of informed nodes provided by the greedy algorithm and the method based on community centrality are statistically similar. Therefore, since the community-based method is computationally faster than the greedy algorithm, it is more suitable to address the influence maximization problem in practice.

\section{Conclusion}

We have analyzed the role of degree-degree correlation in the influence maximization problem. To simulate the information spreading, we consider the rumor model proposed by Maki and Thompson~\cite{Maki1973,Vega-Oliveros2019a}, which is more suitable to represent the information dynamics in social networks~\cite{Pastor-Satorras2015}. We have proposed a method to maximize the influence transmission based on network community organization. This method has been analyzed by performing simulations on the top of eight real and ten artificial complex networks. We have verified that our method is statistically similar, in terms of the information reach, to the approach based on a greedy algorithm, which is computationally expensive. Thus, our results suggest that our method is more suitable in practice since it provides similar results as the greedy approach, but it is less time-consuming.

Our analysis can be extended by comparing our method with the exact solution of the optimal set of seed in small networks. We can generate networks with varying level of community organization and determine the set of seed by considering both methods. Then, we can verify for what cases our approach is close to the optimal solution. Our analysis can also be complemented with the consideration of patterns of connections and influence inside networks (e.g.~\cite{Vega-Oliveros2019b, ZHOU201969, Chen2019,Costa09}) to select the set of initial spreaders. Besides, centrality measures like proposed in \cite{Flavio2018,arnaudon2019graph,ghalmane2019centrality,Tixier2019arxiv} have potential for finding the best seeds in the IMP, and deserve future studies.

Another interesting phenomenon is the emerging of a maximum value of influence related to seed size and network structure. The peak seems to be related to both the modularity and degree-degree correlation, in which: the more disassortative and lower modular the network, the peak appears with the best-ranked nodes; and when more disassortative and higher modularity, it occurs in the community seeds. The before indicates that there exists a crossover around $\rho \thicksim 0$. We consider this point deserves more analysis in future works.

The study of weighted~\cite{Barrat04}, multilayer~\cite{Boccaletti2014}, and dynamical networks~\cite{Holme012} is also promising. In all these cases, general methods for community identification in networks are necessary~\cite{Fortunato16}.

\section*{Acknowledgments}

Research carried out using the computational resources of the Center for Mathematical Sciences Applied to Industry (CeMEAI) funded by the grant 2013/07375-0 São Paulo Research Foundation (FAPESP).
D.A.VO. acknowledges CNPq (grant 140688/2013-7) and FAPESP (grants {2016/23698-1} and {2018/24260-5}). F.A.R. acknowledges CNPq (grant 305940/2010-4) and FAPESP (grant 2016/25682-5). L.F.C thanks CNPq (307333/2013-2), FAPESP ({2011/50761-2}), and NAP-PRP-USP for the financial support. This research is also supported by FAPESP (grant 2015/50122-0) and DFG-GRTK (grant 1740/2).

\newpage
\appendix

\section{Results of Assortativity in Artificial Networks}
\label{sec:appendixA}
\setcounter{figure}{0}

\begin{figure*}[!h]
\centering
\subfigure[BA ($\rho=-0.21$, $Q=0.25$)]{\includegraphics[scale=0.23]{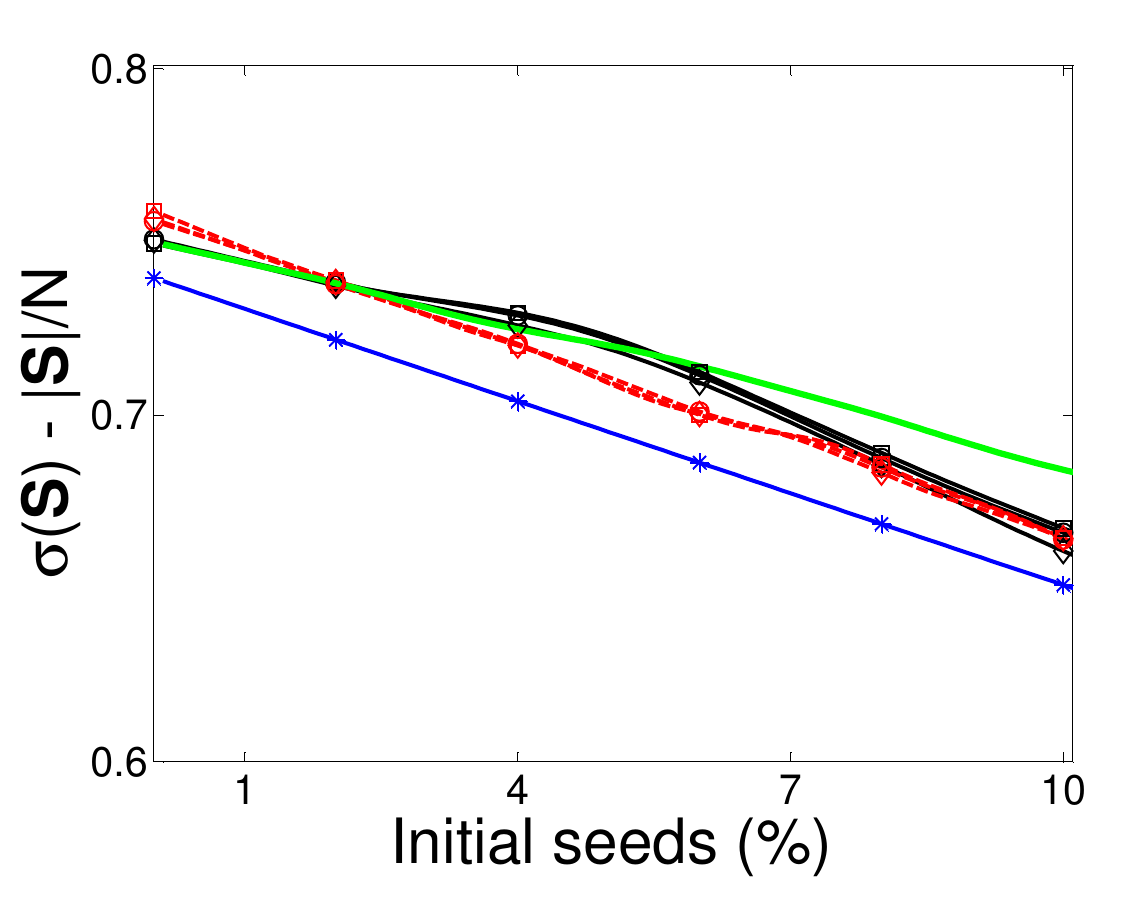}}  
\subfigure[BA ($\rho=0.34$, $Q=0.27$)]{\includegraphics[scale=0.23]{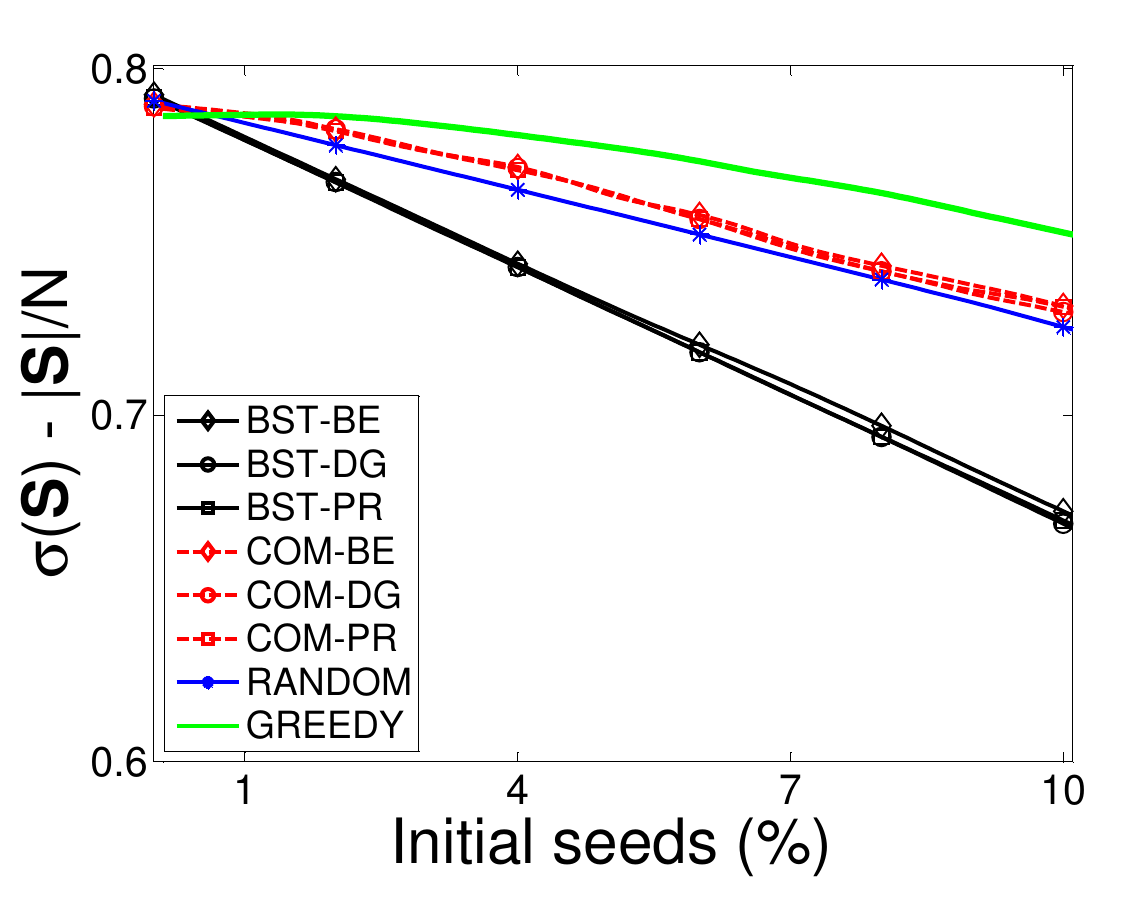}}  
\caption{Impact of degree-degree correlation on the influence maximization problem, when maximizing the relative size of the outbreak in function of the initial seeds (\Sd). The number of seeds varies from two nodes to $10\%N$. For each artificial scale-free network, we calculate the set of initial seeds according to: (BST) the best-ranked nodes of the network; (COM) the most central nodes from communities; (RANDOM) randomly selecting the initial seeds; and (\emph{GREEDY}) the greedy method. The adopted measures are betweenness centrality (\emph{BE}), degree (\emph{DG}) and PageRank (\emph{PR}).}
\label{fig:resArtfAnnex}
\end{figure*}

We have in Fig.~\ref{fig:resArtfAnnex} two networks with similar structural properties, but opposite degree-degree correlation. The analyzed methods produce different results in each case. For instance, in the disassortative network (Fig.~\ref{fig:resArtfAnnex}(a)) the greedy approach has poor performance for lower sizes of {\Sd}, the community seeds reach larger relative outbreak until 2\%N, and the global seeds are a good option as well until 6\%N. On the other hand, in the assortative case (Fig.~\ref{fig:resArtfAnnex}(b)), the greedy approach always achieved the maximum relative outbreak, and the global centrality seeds reached the lowest results, worst than the random selection of seeds.

\section{Spreading parameters evaluation}
\label{sec:appendixB}
\setcounter{figure}{0}

\begin{figure*}[!h]
\centering
  \subfigure[COM-DG]{\label{fig:BA-TPC}\includegraphics[scale=0.23]{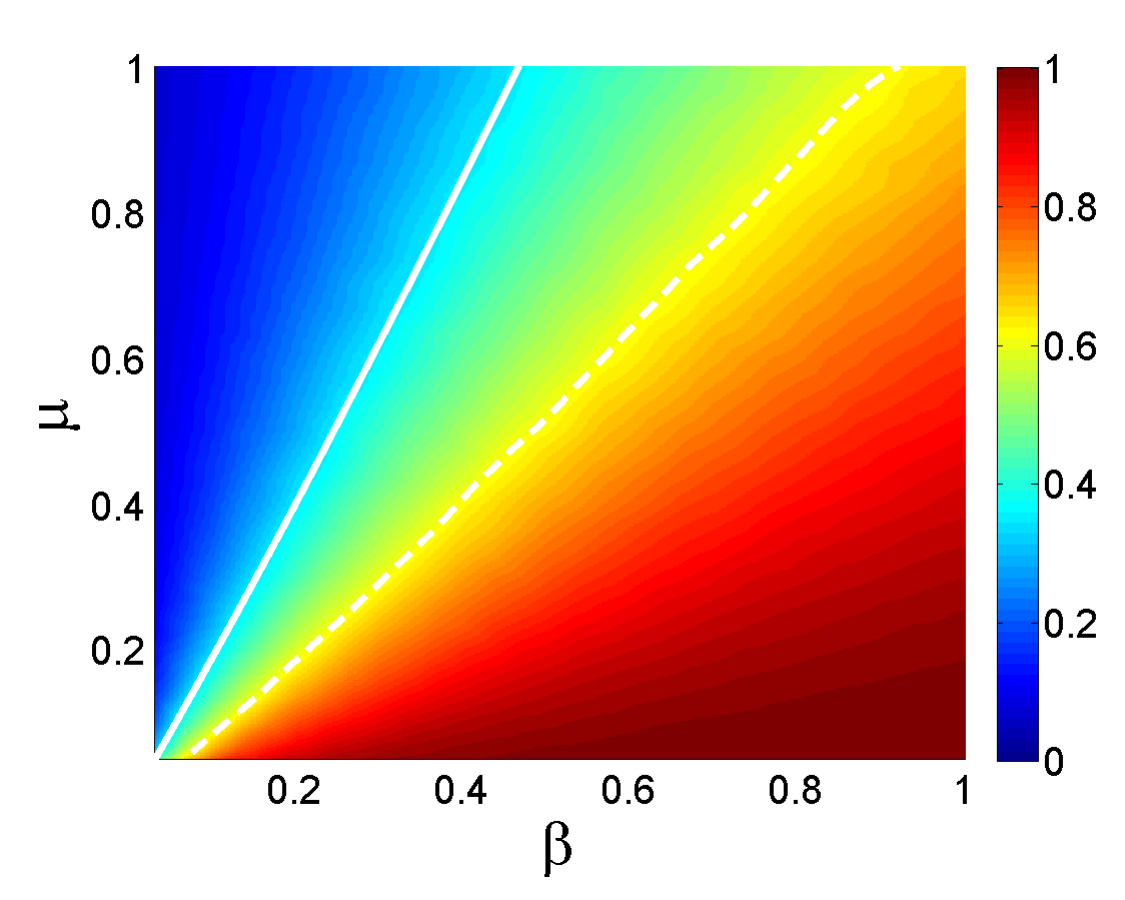}} 
  \subfigure[BST-DG]{\label{fig:BA-TPB}\includegraphics[scale=0.23]{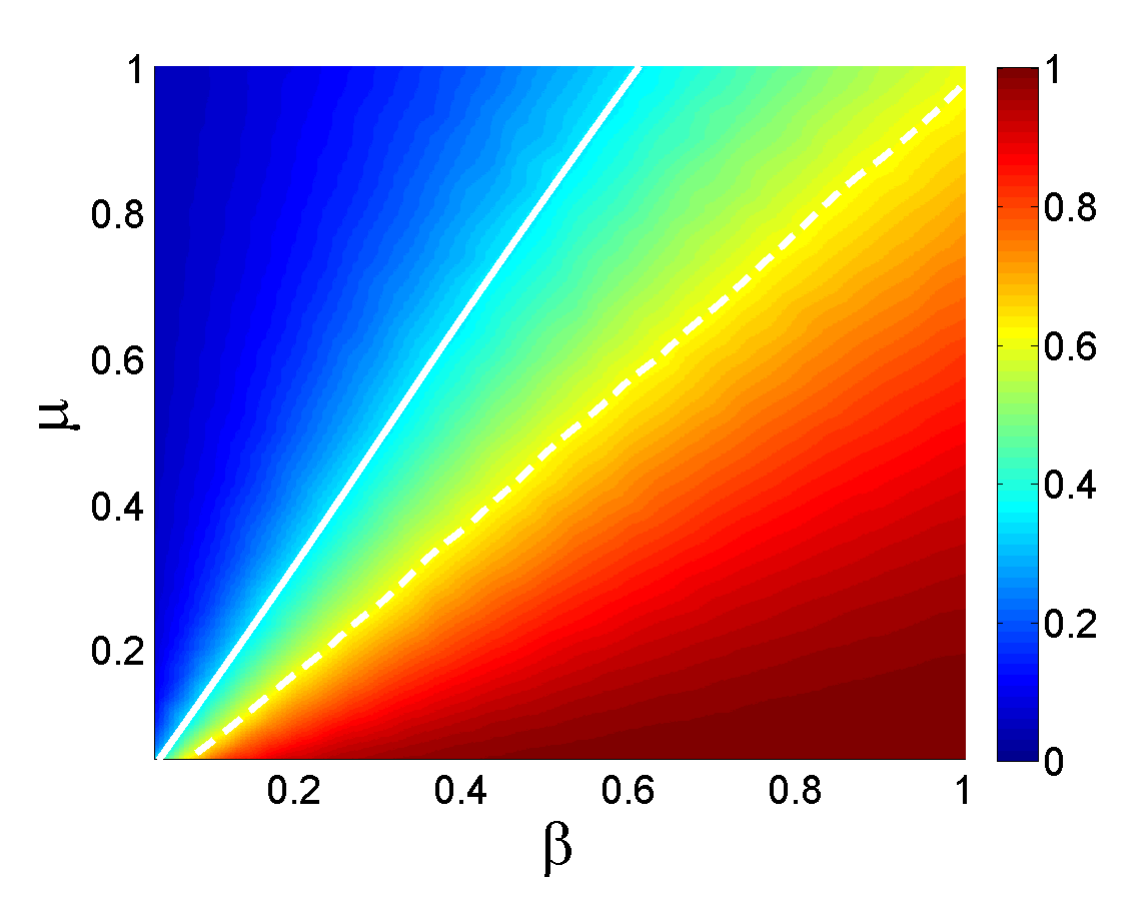}}    
  \subfigure[Random]{\label{fig:BA-TPR}\includegraphics[scale=0.23]{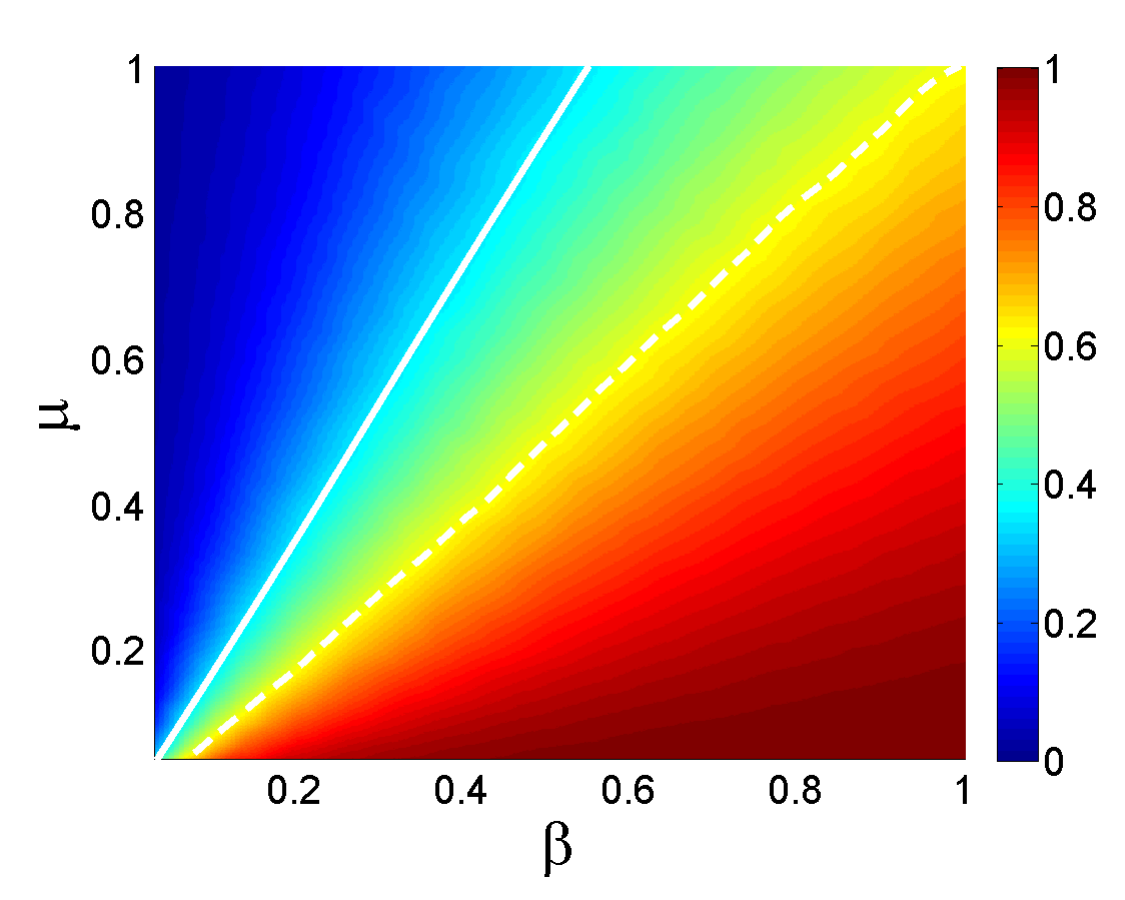}}    
\caption{\label{fig:rumSIRSeeds} \small (color online) Parameter evaluation of the final fraction of informed individuals in the influence maximization problem on a Barab{\'a}si-Albert (BA) network. The BA is characterized with size $N = 1000$, average degree $\left\langle k\right\rangle = 8$, degree-degree correlation  $\rho = -0.04$, and modularity $Q = 0.31$. The final fraction of stiflers are shown in the color bar. The number of seeds is $\eta = 4\%N$, and obtained by the methods: (COM-DG) the local hubs from each community; (BST-DG) the best-ranked nodes with larger degree of the network; and (RANDOM) randomly selecting the initial seeds. Solid white lines show the $\beta$ and $\mu$ combinations that reach a final fraction of $0.35$ informed individuals. Dotted white lines show the combinations that reach a final fraction $0.6$ of stiflers.}
\end{figure*}

We evaluated the relation between the spreading parameters and the selection of the initial spreaders in an artificial scale-free network. In Fig.~\ref{fig:rumSIRSeeds}, we show the simulations adopting the degree centrality and a fixed value of $\eta = 4\%N$ initial spreaders. 

The maximum influence spreading $\Inf(\Sd)$ is affected according to the respective method (Fig.~\ref{fig:rumSIRSeeds}). The solid and dotted white curves represent the combination of $\beta$ and $\mu$ parameters that reach the final fractions of $0.35$ and $0.6$ stiflers, respectively. We observe that these curves show a well defined linear pattern, which means, the results are stable and equivalent values of $\lambda = \beta/\mu$ obtain similar $\Inf(\Sd)$ results.

The white lines of the community centrality selection (Fig.~\ref{fig:rumSIRSeeds}(a)) present higher slopes than the Best-ranked (Fig.~\ref{fig:rumSIRSeeds}(b)) and Random (Fig.~\ref{fig:rumSIRSeeds}(c)) seed selection. This result means that the community selection maximized the spreading influence on the network in the function of {\Sd} more than the other two methods. Additionally, the best-ranked method has the lowest slope in the white lines.

\bibliographystyle{model1-num-names}

\end{document}